\documentclass[adp,a4paper,fleqn]{w-art}
\usepackage{times,cite,w-thm}

\theoremstyle{plain}

\theoremstyle{definition}

\usepackage[]{graphicx}
\usepackage[T1]{fontenc}

\newcommand{\eq}{\begin{equation}}
\newcommand{\feq}{\end{equation}}
\newcommand{\eqn}{\begin{eqnarray}}
\newcommand{\feqn}{\end{eqnarray}}
\newcommand{\arr}{\begin{eqnarray*}}
\newcommand{\farr}{\end{eqnarray*}}

\newcommand{\RR}{{\cal R}}

\font\mybb=msbm10 at 12pt
\def\bb#1{\hbox{\mybb#1}}

\def\bR {\bb{R}}

\numberwithin{equation}{section}

\begin{document}

\DOIsuffix{theDOIsuffix} %%

\Volume{} \Month{} \Year{} %%
\pagespan{1}{} %%
\Receiveddate{27 May 2008} \Reviseddate{18 July 2008} \Accepteddate{XX July 2008 by F. W. Hehl}
\Dateposted{XXXX} %%

\title[Special Relativity in the 21$^{\rm st}$ century]{Special Relativity in the 21$^{\rm st}$ century}

\author[S. Cacciatori]{S. Cacciatori\inst{1,}\footnote{E-mail:~\textsf{sergio.cacciatori@uninsubria.it}}}

\author[V. Gorini]{V. Gorini\inst{1,}%
  \footnote{E-mail:~\textsf{vittorio.gorini@uninsubria.it}}}
\address[\inst{1}]{Dipartimento di Fisica e Matematica, Universit\`a dell'Insubria,
via Valleggio 11, 22100 Como, Italy and INFN, Sez. di Milano, Italy}
%%    Information for the third author
\author[A. Kamenshchik]{A. Kamenshchik\inst{2,3,}\footnote{E-mail:~\textsf{alexander.kamenshchik@bo.infn.it}}}
\address[\inst{2}]{Dipartimento di Fisica and INFN, via Irnerio 46, 40126 Bologna, Italy}%%
\address[\inst{3}]{L.~D.~Landau Institute for Theoretical Physics of the RAS, Kosygin street 2, 119334
Moscow, Russia}
\dedicatory{We dedicate this paper to Andrzej Kossakowski on the occasion of his $70^{\rm th}$ birthday.}
\begin{abstract}
This paper, which is meant to be a tribute to Minkowski's geometrical insight, rests on the idea that
the basic observed symmetries of spacetime homogeneity and of isotropy of space, which are displayed by the spacetime manifold in the limiting
situation in which the effects of gravity can be neglected, leads to a formulation of special relativity based on the appearance of two universal
constants: a limiting speed $c$ and a cosmological constant $\Lambda$ which measures a residual curvature of the universe, which is not ascribable
to the distribution of matter-energy. That these constants should exist is an outcome of the underlying symmetries and is confirmed by experiments and
observations, which furnish their actual values. Specifically, it turns out on these foundations that the kinematical group of special relativity
is the de Sitter group $dS(c,\Lambda)=SO(1,4)$. On this basis, we develop at an elementary classical and, hopefully, sufficiently didactical level
the main aspects of the theory of special relativity based on $SO(1,4)$ (de Sitter relativity). As an application, we apply the formalism to an
intrinsic formulation of point particle kinematics describing both inertial motion and particle collisions and decays.

\end{abstract}

\maketitle

\tableofcontents
\section{Introduction}\label{introduction}
It seems appropriate to us, on the occasion of the centennial of Hermann Minkowski's geometrical formulation of special relativity,
which he put forward in his celebrated address ``Raum und Zeit'' delivered on the 21$^{\rm st}$ of September 1908, at the 80$^{\rm
th}$ Assembly of German Natural Scientists and Physicians at Cologne \cite{Mink}, to pay tribute to this great scientist and
mathematician by providing a contribution to the interpretation and further development of his ideas and intuitions
in the light of the progress which has been achieved in our understanding of the universe in the last hundred years.\\
\indent
Special relativity was originally formulated by Einstein in the ``annus mirabilis'' 1905 resting on 1) {\it the principle of
relativity}, stating the equivalence of the class of all inertial observers as regards the description of the laws of nature and on 2)
{\it the principle of constancy of the velocity of light} which states that electromagnetic waves propagate in empty space with a
universal speed $c$ ( $= 2.99 792 458 \times 10^{10} cm\ s^{-1}$) which does not depend on the state of motion of its source and
of the observer who measures this speed \cite{Ein}.\\
\indent
Strictly speaking Einstein's original formulation differs from the above to the extent that he does refer all motions to some
particular inertial observer which is assumed conventionally to be ``stationary''. However, that this observer does not play any
privileged role compared to the other ones is clear from his statement that \guillemotleft the introduction of a ``luminiferous aether'' will
prove to be superfluous inasmuch as the view here to be developed will not require an ``absolutely stationary space'' provided with
special properties, nor assign a velocity-vector to a point of the empty space in which electromagnetic processes take place\guillemotright
\cite{Ein}.\\
\indent
Thus, Einstein's insight irretrievably shattered the notions of absolute space and absolute time which were deeply rooted in the
mind of scientists since the time of Newton.  However, it was soon realized after Einstein's great breakthrough that the principle of
constancy of the velocity of light was not necessary in order to obtain the Lorentz transformations.\\
\indent
Indeed, beginning with a work by Ignatowski \cite{Ignat}, a large amount of literature has since appeared in which the Lorentz
transformations are deduced solely on the basis of the principle of relativity. Precisely, one assumes the existence of a class of
privileged reference frames ($t,x,y,z,$) for the formulation of the laws of nature (the inertial frames). Then one proves that the
set of spacetime transformations $(t,x,y,z) = (x^{\mu};\mu =0,1,2,3) \rightarrow (t',x',y',z') = (x'^{\mu})$ connecting the
coordinates of any two such frames is the Poincar\'e group $P(c)$ corresponding to an unspecified invariant speed $c$, together with
its limits $c \rightarrow \infty$ (the Galilei group $G$) and $c \rightarrow 0$ (the Carroll group $C$) \cite{LevyLeblond}. Among the
relevant papers we mention only a few representative ones \cite{Frank,Lalan,LevyLeblond1,Gorini,Sen,Giulini,Silagadze}. A
common characteristic of all these derivations of the Lorentz transformations and, more generally, of the Poincar\'e group is that
they essentially proceed in two steps. First, one exploits spacetime homogeneity to prove that the transformations $x'^{\mu} =
f^{\mu}(x^0,x^1,x^2,x^3)$ are affine: $x'^{\mu} =\Lambda^{\mu}_{\nu} x^{\nu} + a^{\mu}$. Second, one uses the
isotropy of space and the assumption of noncompactness of boosts to prove that the set of matrices $\Lambda^{\mu}_{\nu}$ represents the Lorentz
group $L(c)$.  It is implicit in the first step of the above procedure, the one leading to affine transformations, that not only one assumes
spacetime to be homogeneous, but also that this basic symmetry of the spacetime manifold is carried by the privileged
class of observers as well. As clearly stated in \cite{LevyLeblond1} (where for simplicity two-dimensional transformations are
considered): \guillemotleft we assume first that spacetime is {\it homogeneous}, in that it has ``everywhere and everytime'' the same properties \guillemotright.
This statement, ``everywhere and everytime'', means that one has already separated space from time thus selecting a reference frame,
which is itself assumed to be homogeneous. And indeed the author continues: \guillemotleft more precisely, the transformation properties of a
spatiotemporal interval ($\Delta x,\Delta t$) depend only on that interval and not on the location of its end points (in the
considered reference frame). In other words, the transformed interval $\Delta x',\Delta t'$ obtained through an inertial
transformation {\it is independent of its end points}\guillemotright . It is precisely this last property which leads to an affine map. However,
a maximally symmetric manifold with constant non-zero curvature is certainly homogeneous (think for example of a two-dimensional
sphere), but no coordinate system on the manifold can itself be homogeneous. Indeed, if it were so, the components of the metric
tensor relative to such coordinate system would be independent of the point of the manifold, which implies zero curvature. This
indicates that, in searching for the possible structure of the kinematical group of nature reflecting the basic symmetries of
spacetime, the principle of relativity itself, at least in its traditional formulation, has to be abandoned: {\it the basic
spacetime symmetries do not imply the existence of a privileged class of global observers}.\\
\indent
Now then, where does Minkowski enters this picture? Though in his Cologne address he does not state matters in the form that we
have outlined here, {\it it was actually he who first abandoned the principle of constancy of the velocity of light}. Indeed, his point
of departure is the observation that the laws of Newtonian mechanics are invariant under two distinct groups of transformations. The
first one, which we denote by $ST$, is the one generated by space rotations, space translations and time translations. As such, it
embodies the fundamental symmetries of space and the homogeneity of time. The second group is the group $B$ of boosts, namely the group
of those transformations which change the state of motion of the underlying system of spatial coordinates by imparting to it any
uniform translationary motion. Minkowski notes that these two groups \guillemotleft lead their lives entirely apart\guillemotright\
and appear to be utterly
heterogeneous, the group $B$, in addition, appearing to be somewhat accidental and less fundamental than $ST$ (this is for
example the attitude taken by Lorentz and Poincar\'e). However, in Minkowski's opinion this separation appears to be artificial
and it is his firm belief that in Nature the two invariances must be intimately interconnected. The obvious way by which this can be
achieved is to abandon the commutativity of boosts by acknowledging that space and time are not separate entities and that boosts do
``rotate'' space and time among themselves in a way analogous to rotations of the space axes. In this way, space and time cease to
exist as separate absolute entities and give way to a new absolute four-dimensional entity which Minkowski calls {\it the world}
(spacetime, in modern terminology). Then, once we have fixed at will a spacetime point as the origin of the $(t,x,y,z)$-coordinates, the
kinematical group of the world stemming from the basic symmetries of boosts and space rotations is the Lorentz group $L(c)$ (denoted
$G_c$ by Minkowski) which is determined up to the unspecified value of the invariant speed $c$. Thus, it is only in the limit
$c\rightarrow \infty$ that space and time split as separate entities and one recovers, as a degenerate limit, the former two groups $ST$
and $B$ of Newtonian mechanics. In this way, the invariant speed $c$ is not introduced in the argument from the outset, as a physical
velocity experimentally found in nature. Instead, its existence, as an unspecified dimensional parameter, stems from the principle of
relativity and it is only {\it a posteriori} that it can be identified with the physical velocity of propagation of light in
empty space, as Minkowski remarks. Thus he arrives at the Lorentz group (and, by adding spacetime translations, at the Poincar\'e
group) by exploiting the principle of relativity alone. He interprets this principle as implying, through a leap of \guillemotleft audacity
on the part of the higher mathematics\guillemotright, that space and time are intimately interconnected in an indivisible four-dimensional
continuum (world) of events (the term employed by Minkowski is ``substantial points''). It is in this spirit that
he proposes to replace the term ``relativity postulate'' (the principle of relativity) with the term
``{\it postulate of the absolute world}''.\\
\indent
However, following Dyson \cite{Dyson}, we are led to observe here that \guillemotleft Minkowski in his 1908 lecture failed to carry his
argument to its logical conclusions\guillemotright. Indeed, the argument that guided Minkowski to drop the commutativity of boosts thereby
allowing him to interconnect boosts and space rotations into the Lorentz group has left spacetime translations out of the picture. On
the other hand, it is only logical that the same argument should be extended to translations, by dropping their
commutativity as well. Indeed, it appears that, even at the beginning of the 20$^{\rm th}$ century, astronomers would not have had any a priori
compelling reason to believe that the commutativity of translations would not eventually break down provided one would look sufficiently
far away in spacetime. The reason why such a possibility did not occur to scientists then is very likely due to the fact that nobody
had yet any clear knowledge at that time of the existence of other galaxies, in the universe, beyond ours. In fact, it was only through
his observations with the 100 inch Hooker telescope at Mount Wilson Observatory, carried out during the years 1923 and 1924, that
Hubble was able to establish beyond doubt that most of the nebulae, earlier observed with less powerful telescopes and thought
to belong to the Milky Way, were not part of our galaxy, being instead themselves galaxies in their own right, beyond our
own. Therefore, since the picture one had of the universe at the time of Minkowski was that of an island galaxy immersed in an
otherwise infinite void, the idea of non commuting space and time translations would have indeed appeared quite unnatural to
everybody, Minkowski himself included. Nevertheless, it is perhaps a bit surprising that no \guillemotleft act of audacity on the part
of higher mathematics\guillemotright\  arose suggesting that translations were not, after all, commutative. In the words of Dyson \cite{Dyson}, an
opportunity was missed by Minkowski and by his fellow mathematicians. Had someone made the further daring leap of
renouncing the commutativity of translations, a quite logical step after the commutativity of boosts had been abandoned, and
acknowledged that Lorentz transformations and spacetime translations should not \guillemotleft lead their lives entirely apart\guillemotright\  but be instead
interconnected, one would have realized that the true kinematical group of nature is not the Poincar\'e group $P(c)$ but the de Sitter
group $dS(c,\Lambda)$, $\Lambda$ being an unspecified ``cosmological constant'' to be determined by observation \cite{Bacry}. Then,
$P(c)$ would arise as a degenerate limit of $dS(c,\Lambda)$ as $\Lambda \rightarrow 0$, in the same way as the Galilei group $G$ is
a degenerate limit of $P(c)$ as $c \rightarrow \infty$. Had the above step been made one would have discovered theoretically
that the universe expands, and that its expansion is accelerated, well before the development of general relativity, Hubble's
observation of the recession of galaxies, and the observation of the apparent luminosity-red shift relation of distant type Ia
supernovae! But an opportunity was missed and, in spite of all what happened in the development of theory and observation in the
meanwhile, it took 60 years after Minkowski's 1908 address before someone was inspired to tackle in all generality the problem of
finding all possible kinematical groups compatible with the basic observational spacetime symmetries of space isotropy, spacetime
homogeneity and boost invariance. This problem was solved in a truly remarkable paper by Bacry and L\'{e}vy-Leblond
which appeared in 1968 \cite{Bacry}. And sure enough the result turned out to be that the
most general such group is $dS(c,\Lambda)$ (de Sitter or anti-de Sitter, depending on the sign of $\Lambda$), depending on a
unspecified invariant velocity $c$ and a likewise unspecified intrinsic spacetime curvature $R = \sqrt{3/ |\Lambda|}$, together
with its possible contractions $c=0,\infty$ and/or $\Lambda=0,\infty$. However, the Bacry and
L\'{e}vy-Leblond paper went largely unnoticed among the community of cosmologists. The reason for this can be perhaps ascribed to the
fact that the paper was written by mathematical physicists and also that the cosmological constant $\Lambda$, which Einstein had
introduced with the purpose of obtaining a static solution of his equations, was dropped after the discovery by Hubble that the universe is
expanding, only for being seriously re-introduced into the picture at the end of last century, when the observation of the
luminosity-redshift relation of distant type Ia supernovae strongly indicated that the expansion of the universe is nowadays accelerated
(and has ever been so since approximately the last five billion years) \cite{cosmic}.\\
\indent
To be sure, even well before the observation of distant supernovae, particle physicists, following an original suggestion by
Zel'dovich \cite{zeld}, had considered the possibility of appearance of a non zero cosmological constant as a form of
gravitationally repulsive ``vacuum energy density'' arising from the vacuum fluctuations of the quantum fields describing the known
elementary particles. However, the value of $\Lambda$ obtained from a heuristic and quite na\"ive estimate of such vacuum energy is 120
orders of magnitude greater than the observed one. Therefore, if $\Lambda$ is quantum mechanical in origin, one is faced with the
shameful problem  of explaining such preposterous discrepancy. Subtle cancellations should occur and one of the possible
applications of supersymmetry is an implementation of such cancellations \cite{cancellations}.\\
\indent
Now, what is the present observational situation? There are several independent observational and numerical  evidences which are all
consistent with each other and which indicate that there is an effect, which is gravitationally repulsive and which is responsible for the
accelerated expansion. In addition to the supernovae observations, this hypothesis is independently supported by the observation of
ripples in the distribution of galaxies, which match the spectrum of baryon acoustic oscillations in the primordial plasma before
decoupling, by the apparent size of cold and hot spots in the cosmic microwave background (CMB), by the reworking of the CMB by the
gravitational fields of cosmic structures (the integrated Sachs-Wolfe effect), by the study of the growth of galaxy clusters
via weak gravitational lensing techniques, and by the observed changes in galaxy population of the universe during the different
cosmic epochs (evolution of galaxy mergers, rate of star formation and galactic metamorphosis \cite{dark-reviews}).\\
\indent
All these observations point to the conclusion that the universe is permeated by some ``exotic'' kind of  gravitationally repulsive
energy density, universally referred to as ``dark energy'', which seems to be distributed uniformly in space (being therefore free of
clumping) which accounts for approximately 75\% of the energy content of the universe, the remaining 25\% being provided by
ordinary matter (baryonic and dark). The equation of state parameter $w$ of dark energy matches within 10\% the value $w=-1$
corresponding to its interpretation as a cosmological constant. Furthermore, this value of $w$ for dark energy does not appear to
have varied significantly in the last seven billion years.\\
\indent
What is the true nature of dark energy? Regarding this question particle physicists and cosmologists are to date groping in pitch
dark. Besides the interpretation of dark energy as a cosmological constant (possibly quantum mechanical in origin) there are several alternative hypotheses regarding the nature of the phenomenon. Among these: some dynamic field varying in space and time (quintessence, \cite{quintessence}); a modification of general relativity operating at large scales \cite{zerbini,barvinsky}; the proposal that the observed acceleration of the universe is ascribable to the relativistic backreaction of cosmological perturbations and does not require the presence of some kind of negative pressure fluid \cite{kolb}; that the expansion of the universe is in fact decelerated and that it is just that gravitational effects arising from the highly uneven distribution of matter in the cosmos affect differently the measurements of space and time from place to place and thereby create the illusion of acceleration (the fractal bubble model, \cite{wiltshire}). With so many theories vying to explain dark energy, more data are needed. And, indeed, a big effort will be carried out in the coming years in terms of  dedicated instrumentation
on earth and in space with the purpose of  better investigating the nature of dark energy, whether it changes in space and time,
whether it is a simple manifestation of the cosmological constant, thus being uniform in space and unchanging in time, or wether it is simply an illusion due to the highly inhomogeneus distribution of matter in the cosmos. Some
representatives of such important programs are the Baryon Oscillation Sky Survey (due to start in 2009), the H-E Telescope
Dark Energy Experiment (planned to begin observations in 2010), several new generation cosmic-microwave-background experiments, both
on the ground and in space (for example the  Planck  satellite, due for launch this year), the Large Synoptic Survey telescope, due to
start operation in 2012, the Dark Energy Space Telescope for the study of supernovae and weak lensing (possibly starting operating in
2014), the Advanced Dark Energy  Physics Telescope, which would study baryon acoustic oscillations and supernovae and which may also
start operating in 2014, and the Joint Dark Energy Mission which may be launched sometime between 2011 and 2017. If we espouse the  point
of view that dark energy is  the cosmological constant, another intriguing aspect of the matter is why its observed value (of the
order of $10^{-56}  cm^{-2}$) is so small and why the corresponding energy density (of the order of $10^{-29} cm\ g^{-3}$) is comparable
to the energy density of ordinary matter (the cosmic coincidence conundrum). It was not so in the distant past, when $\Omega_M$ was
much larger than $\Omega_\Lambda$, and it will presumably be no more so in the far future from now, when we expect to have $\Omega_M \ll
\Omega_\Lambda$ (here, as usual, $\Omega_M$ and $\Omega_\Lambda$ denote respectively the ratios of the densities of ordinary matter
and of dark energy to the critical density). We live during a very special epoch of the history of the universe, an epoch in which
$\Omega_\Lambda$ and $\Omega_M$ are of the same order of magnitude! Perhaps the only way to explain this coincidence is to resort to
some kind of anthropic principle \cite{barrow}, though such an explanation is far from being a satisfactory one.\\
\indent
Quite independently of these problems, on the basis of the analysis displayed in the first part of this introduction, we are
inclined to believe, in Minkowski's spirit, that $\Lambda$ {\it is a true (classical) universal constant}, just as the limiting speed $c$
is. In other words, we adopt the point of view of Bacry and L\'evy-Leblond, who are motivated in finding the most general
kinematical group which is compatible with the seemingly unquestionable fundamental symmetries that nature displays.
Precisely, experiments and observations in all realms of physics indicate that the local laws governing natural phenomena do not
depend on time and on the location in space (spacetime homogeneity), that no direction in space is privileged compared to any other
(isotropy of space), and that the principle of inertia is locally valid (invariance under boosts). The principle incorporating the
combination of these symmetries should then replace the historical principle of relativity and, in Minkowski's spirit, be rightly
termed {\it the principle of the absolute world}.
The most general kinematical group compatible with this principle is $dS(c,\Lambda)$, and the actual values of $c$ and $\Lambda$ must be
determined by observation.\footnote{In sec. \ref{sec:sasha} we show that this result follows essentially from spacetime homogeneity
and isotropy of space alone, the principle of inertia being actually a consequence of these properties.}
And, sure enough, it turns out that both $c$ and $\Lambda$ have well defined nonzero finite values (in
particular, $\Lambda$ is positive). It would be surprising if it were otherwise. This means that the background
arena for all natural phenomena, once all physical matter-energy has been ideally removed, is the (maximally symmetric) de Sitter
spacetime $dS(1,3)$ (whose radius $R$ is related to $\Lambda$ by the equation $R =\sqrt {3/\Lambda}$) in place of the familiar flat
Minkowski spacetime $M(1,3)$, which can be recovered only in the limit $\Lambda \rightarrow 0$. There is of course no compelling
reason to exclude that part of the measured value of the cosmological constant may be indeed due to some form of
gravitationally repulsive stuff whose physical origin still escapes our grasp, so that theoretical and experimental investigations
regarding the nature of this substance are certainly justified. However, we maintain that  at least a residual part of $\Lambda$ has
indeed to be regarded as a fundamental constant of nature.\\
\indent
Thus, in the approximation in which the warping effects of gravitation on the geometry of spacetime can, at least locally, be
neglected, the existence of a residual curvature leads naturally to the problem of reformulating the theory of special relativity as a
de Sitter relativity, based on $dS(c,\Lambda)$ invariance, in place of the customary Minkowskian relativity, based on Poincar\'e
invariance. The formulation of a de Sitter relativity presents nontrivial complications compared to the Minkowski case. Indeed, in
Minkowski space there is a class of privileged reference frames (coordinate systems), the inertial ones relative to
which the laws of nature have a particularly simple form which is form invariant under the transformations of the underlying
kinematical group (Poincar\'e invariance). Instead, due to the presence of curvature, no such class of privileged coordinate
systems exists in the de Sitter manifold, in spite of the fact that the latter, like Minkowski space, is maximally symmetric. Therefore,
in $dS(1,3)$ any coordinate patch is in principle equally suitable for the formulation of a particular law, which will thus acquire a
generally covariant form, and the notion of observer becomes a strictly local one. To overcome this difficulty it is important to
investigate whether and to what extent meaningful physical quantities and properties can be expressed in an intrinsic form,
independent of the choice of any particular coordinate system. Besides introducing and developing the general setting, the contribution of this
paper, which is kept at the purely classical level, consists in carrying out this program for the description of the free motion of
classical particles and of particle collisions in terms of an intrinsic characterization of the associated conservation laws.\\
\indent
Finally, we would like to make the following remark. In the Galilei group both translations and boosts form Abelian subgroups. The
transition from nonrelativistic mechanics to Einstein's special relativity, namely the transition from the Galilei group to that of
Poincar\'e arises due to the noncommutativity of boosts, while translations remain commutative. If one had decided instead to drop
the commutativity of translations while preserving that of boosts, one would arrive at the Newton-Hooke group \cite{Silagadze,Bacry}
which corresponds to a world without limiting velocity, but with a constant spatial curvature.

%%%%%%%%%%%%%%%%%%%%%%%%%%%%%%%%%%%%%%%%%%%%%%%%%%%%%%%%%%%%%%%
\section{Kinematical algebras}\label{algebre}
We make no prior assumptions on spacetime $W$ except that it is a four-dimensional continuum which, according to the principle of the
absolute world, should embody the properties of homogeneity of space and time, of isotropy of space, and of invariance under boosts. This
leads us to assume that its kinematical group is a ten-dimensional Lie group. We follow the treatment of reference \cite{Bacry}.
However, we do not require space reflection and time reversal to be automorphisms of the Lie algebra of the group and exploit only the
transformation properties of the generators under space reflection. We denote by $J_i, P_i, H$ and $K_i$ ($i=1,2,3$) respectively the
generators of space rotations, space translations, time translation and boosts. They form a basis for the Lie algebra of the group. The
isotropy of space implies that {\bf {J}, {P}} and {\bf K} are vectors and that $H$ is a scalar so that the following Lie brackets
should hold
\begin{eqnarray}
&& \left[J_i , H \right] =0 \ ,\quad  \left[J_i , P_j \right] =\epsilon_{ijl} P_l \ ,
\quad \left[J_i , J_j \right] =\epsilon_{ijl} J_l \ , \quad \left[J_i , K_j \right] =\epsilon_{ijl} K_l \label{cinque} \ .
\end{eqnarray}
The remaining Lie brackets have the general form:
\begin{eqnarray}
&& \left[H , P_i \right] =\omega_i H +\gamma_{il} P_l +\varepsilon_{il} J_l +\alpha_{il} K_l \ , \cr
&& \left[H , K_i \right] =\chi_i H +\lambda_{il} P_l +\zeta_{il} J_l +\eta_{il} K_l \ , \cr
&& \left[P_i , P_j \right] =\iota_{ij} H + \varphi_{ijl} P_l +\beta_{ijl} J_l +\psi_{ijl} K_l \ , \label{10} \\
&& \left[K_i , K_j \right] =\xi_{ij} H + \nu_{ijl} P_l +\mu_{ijl} J_l +\upsilon_{ijl} K_l \ ,  \cr
&& \left[P_i , K_j \right] =\rho_{ij} H + \pi_{ijl} P_l+\sigma_{ijl} J_l +\tau_{ijl} K_l \ . \nonumber
\end{eqnarray}
However, these brackets are constrained by the Jacobi identities on the triples
\begin{eqnarray}
&& [{\bf{J}}H{\bf P}]\ , \qquad [{\bf J}H{\bf K}]\ , \qquad [{\bf
JPP}]\ , \qquad [{\bf JKK}]\ , \qquad [{\bf JPK}]\ , \qquad [H{\bf PP}]\ , \cr
&&  [H{\bf KK}]\ , \qquad [H{\bf PK}]\ , \qquad [{\bf PPK}]\ , \qquad [{\bf PKK}]\ , \qquad [{\bf PPP}]\ , \qquad
[{\bf KKK}] \ . \label{triples}
\end{eqnarray}
Among these identities, imposing those which contain {\bf J}, we find that the structure constants must be rotationally invariant tensors
so that
\begin{eqnarray}
&&\!\!\!\!\!\! \omega_i \!=\!\chi_i \!=\!0 \ , \quad \iota_{ij}\!=\!\xi_{ij} \!=\!0 \ , \quad \rho_{ij}\!=\!\rho \delta_{ij} \ , \quad \gamma_{il} \!=\!\gamma \delta_{il} \ ,
   \quad \varepsilon_{il}\!=\!\varepsilon \delta_{il} \ , \quad \alpha_{il} \!=\!\alpha \delta_{il} \ , \cr
&&\!\!\!\!\!\! \lambda_{il} \!=\!\lambda \delta_{il} \ , \quad \zeta_{il} \!=\!\zeta \delta_{il} \ ,
   \quad \eta_{il} \!=\!\eta \delta_{il} \ , \quad \varphi_{ijl} \!=\!\varphi \epsilon_{ijl} \ , \quad
\beta_{ijl}\!=\!\beta \epsilon_{ijl} \ , \quad  \psi_{ijl} \!=\!\psi \epsilon_{ijl} \ ,  \\
&&\!\!\!\!\!\! \nu_{ijl} \!=\!\nu \epsilon_{ijl} \ , \quad \mu_{ijl} \!=\!\mu \epsilon_{ijl} \ , \quad \upsilon_{ijl}
\!=\!\upsilon \epsilon_{ijl} \ ,\quad \pi_{ijl} \!=\!\pi \epsilon_{ijl} \ , \quad \sigma_{ijl} \!=\!\sigma \epsilon_{ijl} \ ,
\quad \tau_{ijl} \!=\!\tau    \epsilon_{ijl} \ , \nonumber
\end{eqnarray}
and the Lie brackets (\ref{10}) become
\begin{eqnarray}
&& \left[H , P_i \right] =\gamma P_i +\varepsilon J_i +\alpha K_i \ , \quad\ \left[H , K_i \right] =\lambda P_i +\zeta J_i +\eta K_i, \cr
&& \left[P_i , P_j \right] =\varphi \epsilon_{ijl} P_l +\beta \epsilon_{ijl} J_l +\psi \epsilon_{ijl} K_l, \quad\
\left[K_i , K_j \right] =\nu \epsilon_{ijl} P_l +\mu \epsilon_{ijl} J_l +\upsilon \epsilon_{ijl} K_l, \label{22} \\
&& \left[P_i , K_j \right] =\rho \delta_{ij} H + \pi \epsilon_{ijl} P_l +\sigma \epsilon_{ijl} J_l +\tau \epsilon_{ijl} K_l.\nonumber
\end{eqnarray}
By further imposing the transformation properties of {\bf P, J} and {\bf K} under space reflection, namely
\begin{eqnarray}
\Pi : ({\bf P,J,K}) \longrightarrow (-{\bf P}, {\bf J}, -{\bf K})\ ,
\end{eqnarray}
gives $\varphi =\psi=\nu=\upsilon=0$ so that (\ref{22})
simplifies to
\begin{eqnarray}
&& \left[H , P_i \right] =\gamma P_i +\varepsilon J_i +\alpha K_i \ , \quad\ \left[H , K_i \right] =\lambda P_i +\zeta J_i +\eta K_i,
\quad\ \left[P_i , P_j \right] =\beta \epsilon_{ijl} J_l, \cr
&& \left[K_i , K_j \right] =\mu \epsilon_{ijl} J_l, \qquad
\left[P_i , K_j \right] =\rho \delta_{ij} H + \pi \epsilon_{ijl} P_l +\sigma \epsilon_{ijl} J_l +\tau \epsilon_{ijl} K_l.\label{28}
\end{eqnarray}
Note that we do not require $H$ to be invariant under space reflection in accordance with the fact that parity is not a symmetry
of nature.\\
\indent
Now we impose on the Lie brackets (\ref{28}) the Jacobi identities for the remaining triples (\ref{triples}). For the
triples $[{\bf PPP}]$ and $[{\bf KKK}]$ these identities are automatically satisfied, whereas for the triples $[H{\bf PP}]$,
$[H{\bf KK}]$, $[H{\bf PK}]$, $[{\bf PPK}]$ and $[{\bf PKK}]$ they imply the relations
\begin{eqnarray}
&&\!\!\!\!\!\!\!\!\!\!\!\! \varepsilon +\alpha \pi =0 \ ,\quad\ \gamma \beta +\alpha \sigma =0 \ , \quad\ \alpha \tau =0 \ ,\quad\ \lambda \pi =0 \ , \quad\
\eta \mu +\lambda \sigma =0 \ , \quad\ \zeta +\lambda \mu =0 \ ,\cr
&&\!\!\!\!\!\!\!\!\!\!\!\! \tau \lambda -\eta \pi -\zeta =0 \ , \quad\ \pi \varepsilon +\tau \zeta - \lambda \beta -\alpha \mu -(\eta +\gamma)\sigma =0 \ , \quad\
\pi \alpha -\gamma \tau -\varepsilon=0 \ , \cr
&&\!\!\!\!\!\!\!\!\!\!\!\! (\gamma +\eta)\rho =0 \ ,\quad\ \rho \gamma +\sigma +\tau \pi =0 \ , \quad\ \rho \epsilon +\pi \beta +\tau \sigma =0 \ , \quad\
\rho \alpha +\tau^2 -\beta =0 \ , \label{pappardella} \\
&&\!\!\!\!\!\!\!\!\!\!\!\! \tau \rho =0 \ ,\quad\ \mu +\lambda \rho -\pi^2 =0 \ , \quad\ \rho \zeta -\pi \sigma -\tau \mu =0 \ , \quad\
\rho \eta -\pi \tau -\sigma =0 \ , \quad\ \pi \rho =0 \nonumber \ .
\end{eqnarray}
The next step consists in finding all solutions of these equations, compatible with the condition that boosts should be noncompact. We
do this under the assumption
\begin{eqnarray}
\rho\neq 0\ , \qquad \alpha\neq 0\ , \qquad \lambda \neq 0\
,\label{generic}
\end{eqnarray}
which is the generic case. All remaining cases can be obtained as suitable limits of (\ref{generic}). Collecting together
(\ref{cinque}) with the simplified form that (\ref{28}) takes after imposing (\ref{generic}) on relations (\ref{pappardella}), we get
\begin{eqnarray}
&& \left[J_i , H \right] =0 \ , \quad\ \left[J_i , P_j \right] =\epsilon_{ijl} P_l \ , \quad\ \left[J_i , J_j \right] =\epsilon_{ijl} J_l \ ,
\quad\ \left[J_i , K_j \right] =\epsilon_{ijl} K_l \ , \cr
&& \left[H , P_i \right] =\gamma P_i +\alpha K_i \ , \quad\ \left[H , K_i \right] =\lambda P_i -\gamma K_i \ , \quad\
\left[P_i , P_j \right] =\alpha \rho \epsilon_{ijl} J_l \ , \label{Icaso}  \\
&& \left[K_i , K_j \right] =-\lambda \rho  \epsilon_{ijl} J_l  \ ,  \quad\ \left[P_i , K_j \right] =\rho\delta_{ij} H -\gamma \rho
\epsilon_{ijl} J_l \ . \nonumber
\end{eqnarray}
We note that the brackets (\ref{Icaso}) are invariant under the symmetry defined by
\begin{eqnarray}
S:\left\{ {\bf P} \leftrightarrow {\bf K}\ , \ \alpha
\leftrightarrow \lambda\ , \ \gamma \leftrightarrow -\gamma\ , \
 \rho \leftrightarrow -\rho\ \right\} \ ,\label{esse}
\end{eqnarray}
so that we can limit ourselves to consider the case $\rho>0$. Indeed, the case $\rho<0$ can be traced back to $\rho>0$ by the
interchange ${\bf P}\leftrightarrow {\bf K}$. Now consider a nonsingular transformation
\begin{eqnarray}
&& P_i=aP'_i +bK'_i \ , \qquad K_i =cP'_i +dK'_i \ , \label{Nbase}\\
&& ad-bc \neq 0 \ . \label{relation}
\end{eqnarray}
We show that the coefficients $a,b,c$ and $d$ can be chosen such that the relations in (\ref{Icaso}) hold with the replacements
$P_i \rightarrow P'_i, \ K_i\rightarrow K'_i,\ (\alpha,\gamma,\lambda,\rho)\rightarrow (\alpha',\gamma',\lambda',\rho')$ with $\gamma'=0$.
Taking into account (\ref{relation}), this condition leads to the following system of four linear equations in the four
unknowns $a,b,c,d$, in which $\alpha'$ and $\lambda'$ play the role of parameters,
\begin{eqnarray}
&& \gamma a -\lambda' b+\alpha c= \alpha' a-\gamma b-\alpha d=\lambda a-\gamma c -\lambda' d=\lambda b -\alpha' c -\gamma d=0,\label{system}
\end{eqnarray}
plus the equation
\begin{eqnarray}
&& ad-bc=\rho/\rho'.\label{det}
\end{eqnarray}
Once we ensure the existence of solutions of this system by equating its determinant to zero, condition (\ref{det})
fixes the normalization of any solution and, without loss of generality, we are permitted to take $\rho'=\rho$. Indeed, this
choice means that in passing from the pair $({\bf P}, {\bf K})$ to the pair $({\bf P'}, {\bf K'})$, we do not change the time scale.
Therefore, we solve system (\ref{system}) with the condition
\begin{eqnarray}
 ad-bc=1 \ . \label{detuno}
\end{eqnarray}
In order for the determinant of (\ref{system}) to vanish, the parameters $\alpha'$ and $\lambda'$ must be chosen so as to satisfy the condition
$\alpha' \lambda' =\gamma^2 +\alpha \lambda$. With any such choice and dropping primes, the brackets (\ref{Icaso}) now write
\begin{eqnarray}
&& \left[J_i , H \right] =0 \ , \quad\ \left[J_i , P_j \right]=\epsilon_{ijl} P_l \ , \quad\ \left[J_i , J_j \right] =\epsilon_{ijl} J_l \ , \cr
&& \left[J_i ,K_j \right] =\epsilon_{ijl} K_l \ , \quad\ \left[H ,P_i \right] =\alpha K_i \ , \quad\ \left[H , K_i \right] =\lambda P_i \ , \label{IIcaso}\\
&& \left[P_i , P_j \right] =\alpha\rho \epsilon_{ijl} J_l \ , \quad\ \left[K_i , K_j \right] =-\lambda\rho \epsilon_{ijl} J_l \ , \quad\
\left[P_i , K_j \right] =\rho\delta_{ij} H \ . \nonumber
\end{eqnarray}
For a given choice of $\alpha$ and $\lambda$ in (\ref{IIcaso}), the replacement $\alpha\rightarrow
\mu\alpha$ and $\lambda\rightarrow \mu^{-1} \lambda$ ($\mu>0$) amounts to a change of scale in the measurement of spatial distances.
Since now $\gamma=0$, equations (\ref{system}) and (\ref{detuno}) become
\begin{eqnarray}
&& c=(\lambda/\alpha) b,  \quad\ d=a\ , \quad \mbox{  and  } \quad  a^2-\frac {\lambda}{\alpha} b^2=1 . \label{aquadro}
\end{eqnarray}
Therefore, the most general linear transformation (\ref{Nbase}) which leaves the Lie brackets (\ref{IIcaso}) invariant is of the form
\begin{eqnarray}
&& P_i=aP'_i+bK'_i, \quad\ K_i=\frac {\lambda}{\alpha} b P'_i +aK'_i
\label{form}
\end{eqnarray}
where (\ref{aquadro}) holds. Depending on the signs of $\alpha$ and $\lambda$ we have the following table \cite{Bacry}:
\begin{eqnarray}
&& \lambda>0, \quad \alpha>0 \qquad\ SO(1,4): {\mbox{  de Sitter (see next section)}} \label{desitter}\\
&& \lambda>0, \quad \alpha<0 \qquad\ SO(2,3): {\mbox{  anti de Sitter}} \label{antids}\\
&& \lambda<0, \quad \alpha>0 \qquad\ SO(5) \label{so5}\\
&& \lambda<0, \quad \alpha<0 \qquad\ SO(1,4) {\mbox{   with compact
boosts.}}\label{compactboost}
\end{eqnarray}
Cases (\ref{so5}) and (\ref{compactboost}) correspond to compact boosts and have to be ruled out. Thus we are left with de Sitter and
anti de Sitter corresponding to positive and negative cosmological constants, respectively. Since observations point to a positive
cosmological constant, we henceforth select the de Sitter case (\ref{desitter}). With such a choice, (\ref{form}) reads
\begin{eqnarray}
&& P_i=\cosh \beta P'_i+\sqrt {\frac \alpha\lambda} \sinh \beta K'_i, \quad\ K_i=\sqrt {\frac {\lambda}{\alpha}} \sinh \beta P'_i
+\cosh \beta K'_i, \label{riform}
\end{eqnarray}
and the arbitrariness of $\beta$ corresponds to different selections of the origin of spacetime.

%%%%%%%%%%%%%%%%%%%%%%%%%%%%%%%%%%%%%%%%%%%%%%%%%%%%%%%%%%%%%%%%%%%%%%%%%%%%%%%%%%%%%%%%%%%%%%%%%%%%%%%%%
%%%%%%%%%%%%%%%%%%%%%%%%%%%%%%%%%%%%%%%%%%%%%%%%%%%%%%%%%%%%%%%%%%%%%%%%%%%%%%%%%%%%%%%%%%%%%%%%%%%%%%%%%
\section{Spacetime manifold}\label{sec:spacetime}
We explicitly construct the spacetime manifold $W$ by identifying it both geometrically and metrically with the quotient
space $dS(c,\Lambda)/L(c)$, where the de Sitter group $dS(c,\Lambda)$ is the Lorentz group $SO(1,4)$ in five dimensions
and $L(c)=SO(1,3)$ is the familiar Lorentz group in four dimensions. This procedure is equivalent to the identification of $W$ with the de
Sitter manifold $dS(1,3)$ which is the one sheeted four-dimensional hyperboloid of equation
\begin{equation}\label{hyperboloid}
(X^0)^2-(X^1)^2-(X^2)^2-(X^3)^2-(X^4)^2=-R^2
\end{equation}
in five-dimensional Minkowski space $M(1,4)$, with $R=\sqrt{3/\Lambda}$.
The differential and metric structure are the ones inherited from those of $M(1,4)$, of which $dS(1,3)$ is a
submanifold.\\
\indent
First we show that $dS(c,\Lambda)=SO(1,4)$. Indeed, choose in $M(1,4)$ inertial coordinates $\{X^A;\ A=0,1,2,3,4;\ X^0=cT\}$, the
metric being $\eta={\rm diag} \{1,-1,-1,-1,-1\}$. Let $f$ be a function on $M(1,4)$ and $L=\bb{I}+\epsilon M$ ($\epsilon \ll 1$) be
an infinitesimal transformation of $SO(1,4)$. If $X\in M(1,4)$, we have
\begin{equation}
f(X+\epsilon MX)=f(X)-\epsilon (Nf) (X). \label{transform}
\end{equation}
The correspondence $M\mapsto N$ is an isomorphism of (the self-representation of) the Lie algebra $so(1,4)$ of $SO(1,4)$ onto
the same Lie algebra realized in terms of differential operators on the space of functions on $M(1,4)$. Corresponding bases $\{
M^{AB}\}$ and $\{N^{AB}\}$ for these two realizations are respectively
\begin{eqnarray}
(M^{AB})^C_{\ \ D}=\eta^{AC} \delta^B_{\ \ D} -\eta^{BC} \delta^A_{\ \ D}
\qquad\ \mbox{ and } \qquad
N^{AB}=X^A \frac {\partial}{\partial{X_B}}-X^B \frac {\partial}{\partial{X_A}} ,\label{generators}
\end{eqnarray}
and they satisfy the same Lie brackets. For example,
\begin{eqnarray}
[N^{AB}, N^{CD}]=\eta^{AD} N^{BC} +\eta^{BC} N^{AD}-\eta^{AC} N^{BD}-\eta^{BD} N^{AC}\ .\label{commutators}
\end{eqnarray}
Then, setting
\begin{eqnarray}
H=\frac cR N_{04}, \ P_i=\frac 1R N_{i4},\ K_i=\frac 1c N_{0i},\ J_i=\frac 12 \epsilon_{ijk} N^{jk} \ \ (i,j,k=1,2,3),
\label{formule}
\end{eqnarray}
where
\begin{equation}
N_{AB} =\eta_{AC}\eta_{BD} N^{CD},
\end{equation}
the commutators (\ref{commutators}) can be rewritten as
\begin{eqnarray}
&& \left[J_i , H \right] =0 \ , \label{97} \quad\ \left[J_i , P_j \right] =\epsilon_{ijl} P_l \ , \quad\
\left[J_i , J_j \right] =\epsilon_{ijl} J_l \ , \\
&& \left[J_i , K_j \right] =\epsilon_{ijl} K_l \ , \quad\ \left[H , K_i \right] =P_i \ , \quad\
\left[P_i , K_j \right] =\frac 1{c^2} \delta_{ij} H \ , \\
&& \left[H , P_i \right] =\frac {c^2}{R^2} K_i \ , \label{101} \\
&& \left[P_i , P_j \right] =\frac 1{R^2} \epsilon_{ijl} J_l \ , \label{103} \\
&& \left[K_i , K_j \right] =-\frac 1{c^2} \epsilon_{ijl} J_l  \ , \label{104}
\end{eqnarray}
which are equivalent to (\ref{IIcaso}) via the rescalings
\begin{eqnarray}
&& H\longrightarrow \frac c{R \sqrt {\alpha \lambda}} H ,\quad\ P_i\longrightarrow \frac 1{R\sqrt {\alpha\rho}} P_i,\quad\
J_i\longrightarrow J_i\quad\ K_i\longrightarrow \frac 1{c\sqrt {\lambda\rho}} K_i.
\end{eqnarray}
The geometrical meaning of the positive constant $R=\sqrt {3/\Lambda}$ as the ``spacetime radius'' will become manifest shortly.\\
\indent
Next note that the homogeneity of $W$ (namely, the physical equivalence of all spacetime points) implies that it must be a
homogeneous space of the kinematical group $G=dS(c,\Lambda)=SO(1,4)$, namely that $G$ should operate on $W$ transitively.
In other words, for any two points $X,Y\in W$ there
exists at least one element\footnote{The points of $W$ are denoted by the same symbols as the points of
$M(1,4)$ since, as already stated, $W$ will ultimately be identified with de Sitter space $dS(1,3)$, the one sheeted hyperboloid
(\ref{hyperboloid}) with the metric inherited by the flat metric of $M(1,4)$.} $L\in G$ such that $Y=LX$. Denote by $H_X$ the stabilizer (or isotropy group) of
$X$, namely the maximal subgroup of $G$ which leaves $X$ fixed:
$H_X=\{H|H\in G,\ HX=X\}$, and note that the isotropy groups $H_X$ are all isomorphic through the adjoint action by the elements of
$G$: $H_Y=H_{LX}=LH_X L^{-1}$. Furthermore, since $G$ has no nontrivial centre, two distinct isotropy subgroups have only the
identity $\bb {I}$ in common.\\
\indent
To proceed with the construction of $W$, note that the homogeneous spaces of $SO(1,4)$ as acting on $M(1,4)$ are the
de Sitter hyperboloids $dS(1,3)$ given by (\ref{hyperboloid}), for all possible values of $R\geq 0$. Therefore, we can identify $W$ set
theoretically and differentiably with one such hyperboloid $dS(1,3)$ for some given value $R>0$. However, this is only half the story,
since we need to provide this set with a metric having $dS(c,\Lambda)$ as the isometry group. Obviously, this metric is the
restriction to $dS(1,3)$ of the flat metric on $M(1,4)$ but, in order to adhere to the spirit of constructing $W$ from the
kinematical group $G$ alone, we must arrive at this metric from the natural metric properties of $G$. This will be achieved by endowing
$G$ with the natural metric $ds^2$ induced on it by the Killing metric of its Lie algebra and then establishing an isometry of
$dS(1,3)$ with a suitably constructed four-dimensional submanifold $S$ of $G$, endowed with the metric inherited by $ds^2$ and on which
$G$ acts transitively by adjunction as an isometry group. We start by noting that since the kinematical transformations which keep the
origin fixed are space rotations and boosts, formulas (\ref{formule}) imply that we have implicitly chosen the origin in
$dS(1,3)$ to be the point
\begin{equation}
\bar X=\{0,0,0,0,R\}
\end{equation}
and that the connected component of its stabilizer $H_{\bar X}$ is the proper orthochronous Lorentz group $L(c)=SO(1,3)$ generated by
the $J_l$'s and $K_l$'s. However, $H_{\bar X}$ is larger then $SO(1,3)$ since it contains in addition all elements of the form
$KH$, where $K\in SO(1,3)$ and $H$ is the matrix
\begin{eqnarray}
H=\left(
\begin{array}{ccccc}
-1 & 0 & 0 & 0 & 0\\
0 & -1 & 0 & 0 & 0\\
0 & 0 & -1 & 0 & 0\\
0 & 0 & 0 & -1 & 0\\
0 & 0 & 0 & 0 & 1
\end{array}
\right),
\end{eqnarray}
which is the product of space reflection times time reversal. Henceforth, with an abuse of notation, we will denote by $SO(1,3)$
this larger group, in place of its connected component. The matrix elements of $H$ are
\begin{eqnarray}
H^A_{\ \ B}=-\delta^A_{ B} -\frac 2{R^2} \bar X^A \bar X_B =H(\bar X)^A_{\ \ B}.
\end{eqnarray}
Now, since the matrices of $SO(1,4)$ are defined by the condition $\eta_{CD} L^C_{\ \ A} L^D_{\ \ B} =\eta_{AB}$,
which expresses the conservation of the spacetime interval, if $L\bar X =X$, we have
\begin{eqnarray}
H(X)^A_{\ \ B} =: (LHL^{-1})^A_{\ \ B}=-\delta^A_{B} -\frac 2{R^2} X^A X_B. \label{114}
\end{eqnarray}
Thus the correspondence $F: H(X)\longrightarrow X$ establishes a one to one map onto $dS(1,3)$ of the four-dimensional
submanifold $S$ of $G$ whose points are the matrices given by (\ref{114}). Now we see that $S$ intersects each coset $H(X) H_{\bar
X}$ of the stabilizer of the origin at exactly the point $H(X)$ so that we can identify $S$, and therefore $dS(1,3)$, with the quotient
space $SO(1,4)/SO(1,3)$. We now make use of the important fact that the Lie algebra $so(1,4)$ of $SO(1,4)$ is endowed with the natural
metric $\langle\ , \rangle$, the so called invariant Killing metric, which is nondegenerate since $so(1,4)$ is simple. As shown in
appendix A, this metric induces canonically a metric on $SO(1,4)$ for which the operations of adjunction $M\rightarrow LML^{-1}$ are
isometries. We show that this metric, restricted to $S$ and carried over to $dS(1,3)$ by means of the map $F$, is precisely the
restriction to $dS(1,3)$ of the flat metric $\eta$ of $M(1,4)$, up to an inessential normalization factor.\\
The metric $\langle\ , \rangle$ is defined via the adjoint representation as in (\ref{scalar})
but, since $so(1,4)$ is simple, it can be computed in any faithful representation, up to normalization. In particular, in the
self-representation we have
\begin{eqnarray}
\langle M, N\rangle =\kappa {\rm Tr} (MN) =\kappa \sum_{A,B} M^A_{\ \ B} N^B_{\ \ A},\label{traccia}
\end{eqnarray}
where $\kappa$ is a suitable normalization factor. As illustrated in appendix A, the metric $\langle\ , \rangle$ induces an invariant
metric on the whole group which, restricted to $S$, can be transferred on $dS(1,3)$ as follows. From (\ref{114}) we have
\begin{eqnarray}
dH(X)^A_{\ \ B}= -\frac 2{R^2} (X^A dX_B +X_B dX^A)
\end{eqnarray}
and note that $\left. X_c dX^c \right|_{X^2=-R^2}=0$. Then the metric induced on $dS(1,3)$ via (\ref{traccia}) is
\begin{eqnarray}
&&\!\!\!\!\!\! ds^2= \langle dH(X), dH(X)\rangle =\kappa dH(X)^A_{\ \ B} dH(X)^B_{\ \ A}=\left. -\frac {4\kappa}{R^2} \eta_{AB} dX^A
dX^B\right|_{X^2=-R^2}.
\end{eqnarray}
Choosing the normalization $\kappa=-\frac {R^2}4$, we obtain the expected result: the spacetime manifold $W$ realized as the quotient
space $SO(1,4)/SO(1,3)$ of the de Sitter group by the isotropy group of the origin. It is endowed with the invariant metric induced on $W$
by the Killing metric on $SO(1,4)$ and can be realized as the de Sitter space dS(1,3), namely as the hyperboloid (\ref{hyperboloid})
embedded isometrically in $M(1,4)$ (i.e. with the metric induced by the flat metric of $M(1,4)$). We may call $dS(1,3)$ the {\it
Minkowski hyperbolic spacetime} or the {\it Minkowski hyperboloid}.
%%%%%%%%%%%%%%%%%%%%%%%%%%%%%%%%%%%%%%%%%%%%%%%%%%%%%%%%%%%%%%%%%%%%%%%%%%%%%%%%%%%%%%%%%%%%%%%%%%%%%%%%%
%%%%%%%%%%%%%%%%%%%%%%%%%%%%%%%%%%%%%%%%%%%%%%%%%%%%%%%%%%%%%%%%%%%%%%%%%%%%%%%%%%%%%%%%%%%%%%%%%%%%%%%%%
%%%%%%%%%%%%%%%%%%%%%%%%%%%%%%%%%%%%%%%%%%%%%%%%%%%%%%%%%%%%%%%%%%%%%%%%%%%%%%%%%%%%%%%%%%%%%%%%%%%%%%%%%
\section{Causal structure and coordinate systems}\label{ssec:causal}
In the dimensional choice (\ref{formule}) of the generators of the de Sitter group there appear explicitly the two
constants $c$ and $R$, having respectively the dimensions of a velocity and of a length (the inertial coordinates $X^A$ in $M(1,4)$
have the dimension of a length, whereas $T=X^0/c$ has the dimension of a time). Due to the physical meaning of the generators and to the
structure of their Lie brackets (\ref{97})-(\ref{104}), $c$ represents the limiting speed of propagation of signals in vacuo w.r.t.
any freely falling (geodesic) local observer, while $R$ is the ``radius'' of the Minkowski hyperboloid which is related to its
constant negative Gaussian curvature $K$ by the relation $K=-1/R^2$. As is well known \cite{desitter}, the Lorentzian manifold $dS(1,3)$ is the
maximally symmetric solution of the vacuum Einstein's field equations $R_{\mu\nu}=\Lambda g_{\mu\nu}$ in the presence of a
cosmological constant $\Lambda$, which is related to $R$ by the equation $R=\sqrt {3/\Lambda}$ (or, alternatively, to the Gaussian
curvature by $K=-\Lambda/3$). As indicated in the introduction, the observational values of $c$ and $\Lambda$ are $c=2.99792458\times
10^{10} cm\ s^{-1}$ and $\Lambda\simeq 10^{-56} cm^{-2}$ (corresponding to $R\simeq 10^{28} cm$). Whereas the value of $c$ is
known with great precision, the one of $\Lambda$ is affected by an uncertainty of the order of a few
percent, due to the scanty data available so far (compare the introduction).\\
\indent
The noncommutativity of spacetime translations is expressed by the brackets (\ref{101}) and (\ref{103}). Precisely, the geometrical
meaning of (\ref{103}) is the following. Take a tangent vector to the Minkowski hyperboloid at some point $X$ and displace it parallelly
to itself by an infinitesimal amount first along a space direction $j$, then along a second space direction $i$, and then reverse the
operation coming back to the original point. As a result of this displacement, one ends up with a final vector at $X$ which differs
from the original one by a rotation in the $i$--$j$ plane by a second order angle\footnote{In order to
visualize the geometrical meaning of this fact, it is particularly useful to think of the analogue of this operation performed on the
surface of a sphere, the two directions $i$ and $j$ being for example those of the familiar polar angles $\theta$ and $\phi$.}
proportional to $1/R^2$. Due
to the isotropy of space and to the homogeneity of spacetime, this rotation, which depends on the infinitesimal displacement along the
$i$ and $j$ directions, does not depend on the orientation of these directions in space and on the choice of the point $X$. The
geometrical meaning of the Lie bracket (\ref{101}) is similar, with the infinitesimal displacement along the second space axis replaced
by an infinitesimal displacement in time: upon returning to the original point, the result of the parallel displacement of the vector
is now a second order ``rotation'' in the $i$--time plane, namely a second order boost in the $i$--direction. Both these results are
particular cases of the general formula, proved in appendix B and which applies to any metric manifold, according to which the outcome
of the parallel displacement of a vector along a small closed loop $\gamma$ is expressed in terms of the area bounded by the loop and
of the curvature tensor, the formula being actually interpretable as the definition of the curvature tensor itself.\\
\indent
As regards the noncommutativity of boosts, of which Minkowski was an upholder and which is expressed by (\ref{104}), it has an
interpretation similar to the one of equation (\ref{103}) but referred here to displacements in the space of velocities:
performing an infinitesimal boost in the direction $j$, followed by another such boost in the direction $i$, and then reversing the
order of these operations also results in an infinitesimal rotation in the $i$--$j$ plane (Thomas precession \cite{Thomas}).\\
\indent
The metric
\begin{eqnarray}
\left. ds^2=\eta_{AB} dX^A dX^B \right|_{X^2=-R^2}\label{122}
\end{eqnarray}
of the Minkowski hyperboloid captures the intrinsic geometry of the latter, independently of its embedding in $M(1,4)$. However, this
embedding is very useful to determine all the main properties of $dS(1,3)$, independently from the choice of local coordinates, in
the same way as the main properties of an Euclidean sphere $S^2$ can be deduced from its natural embedding in $\bb {R}^3$. In particular,
in view of the applications expounded in section \ref{sec:kinematics}, we are interested in the nature of geodesics.
We know that geodesic lines in $S^2$ are great circles, and can be obtained by intersecting the sphere with planes in $\bb {R}^3$ passing
through the center of the sphere itself. Similarly, all geodesic curves on $dS(1,3)$ can be obtained by intersecting it with two-planes
in $M(1,4)$ passing through the origin (the center of the hyperboloid). This also permits to describe globally the causal structure of the
Minkowski hyperboloid: the geodesics will by timelike, lightlike or spacelike depending on the inclination relative to the $X^0$ axis of
the corresponding two-planes (see figure 1). This can be seen by introducing the forward lightcone in $M(1,4)$ (the asymptotic cone)
\begin{figure}[h]
\begin{center}
\includegraphics[height=5.8cm]{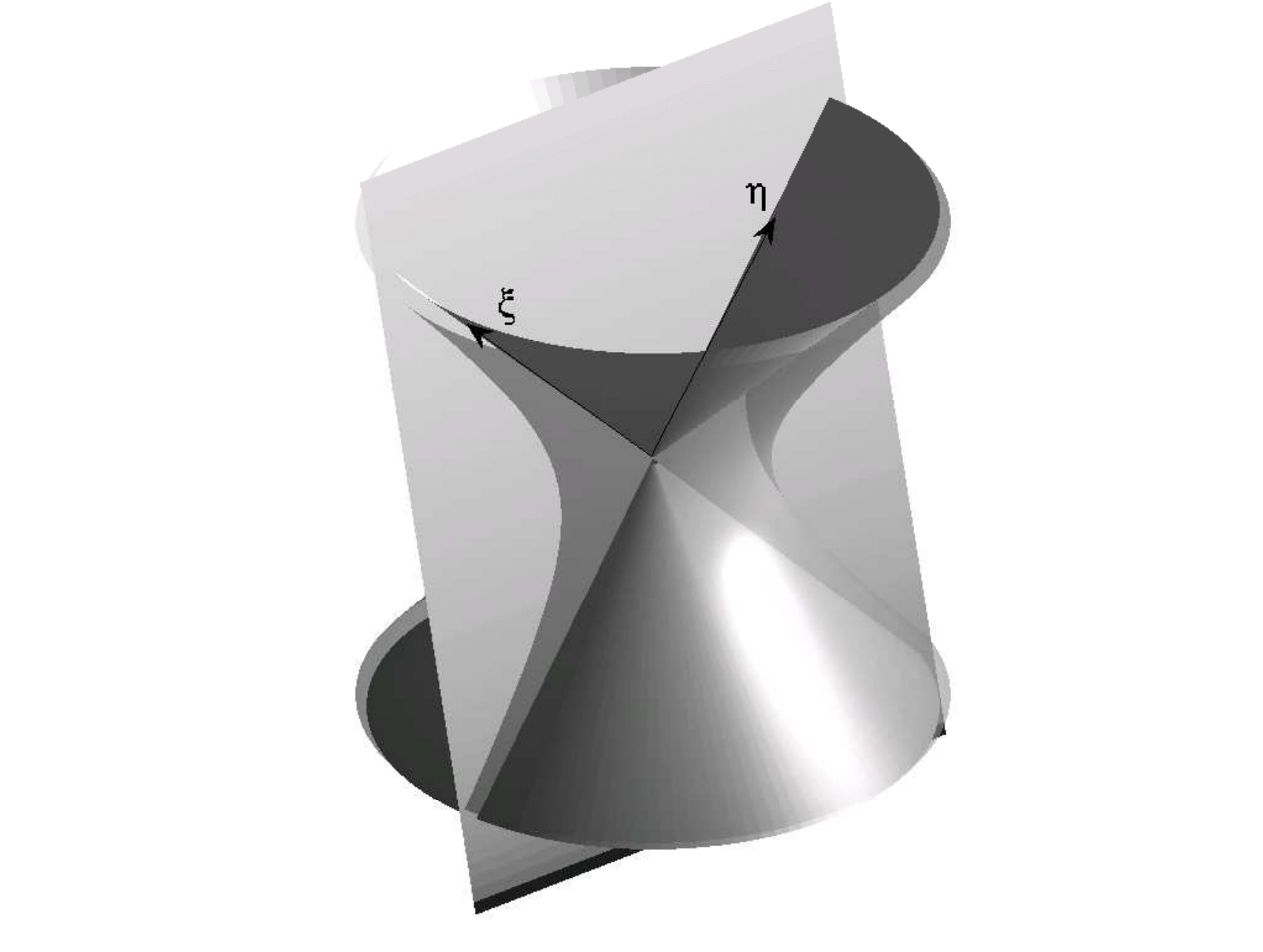}~\includegraphics[height=5.8cm]{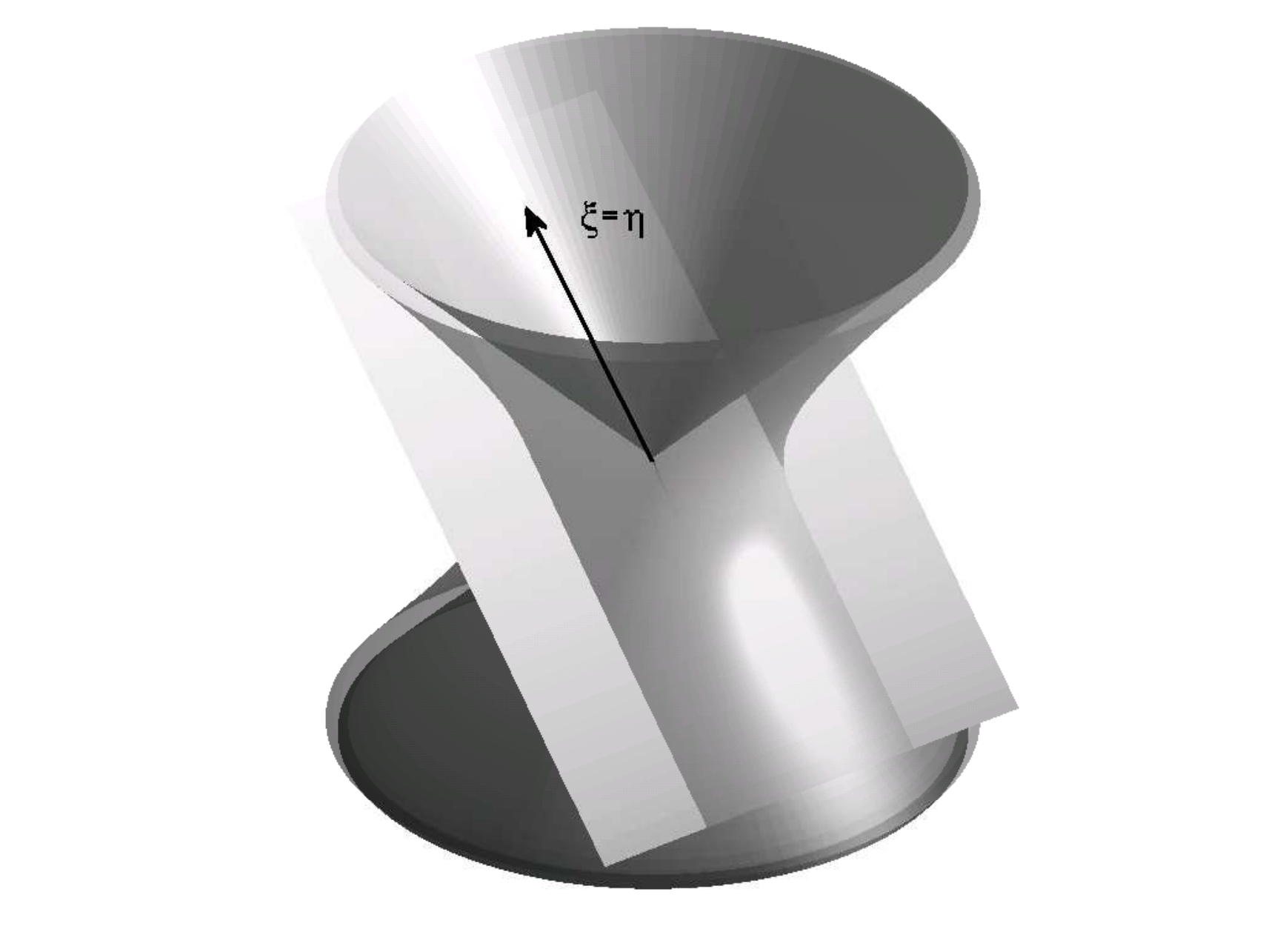} \includegraphics[height=5.8cm]{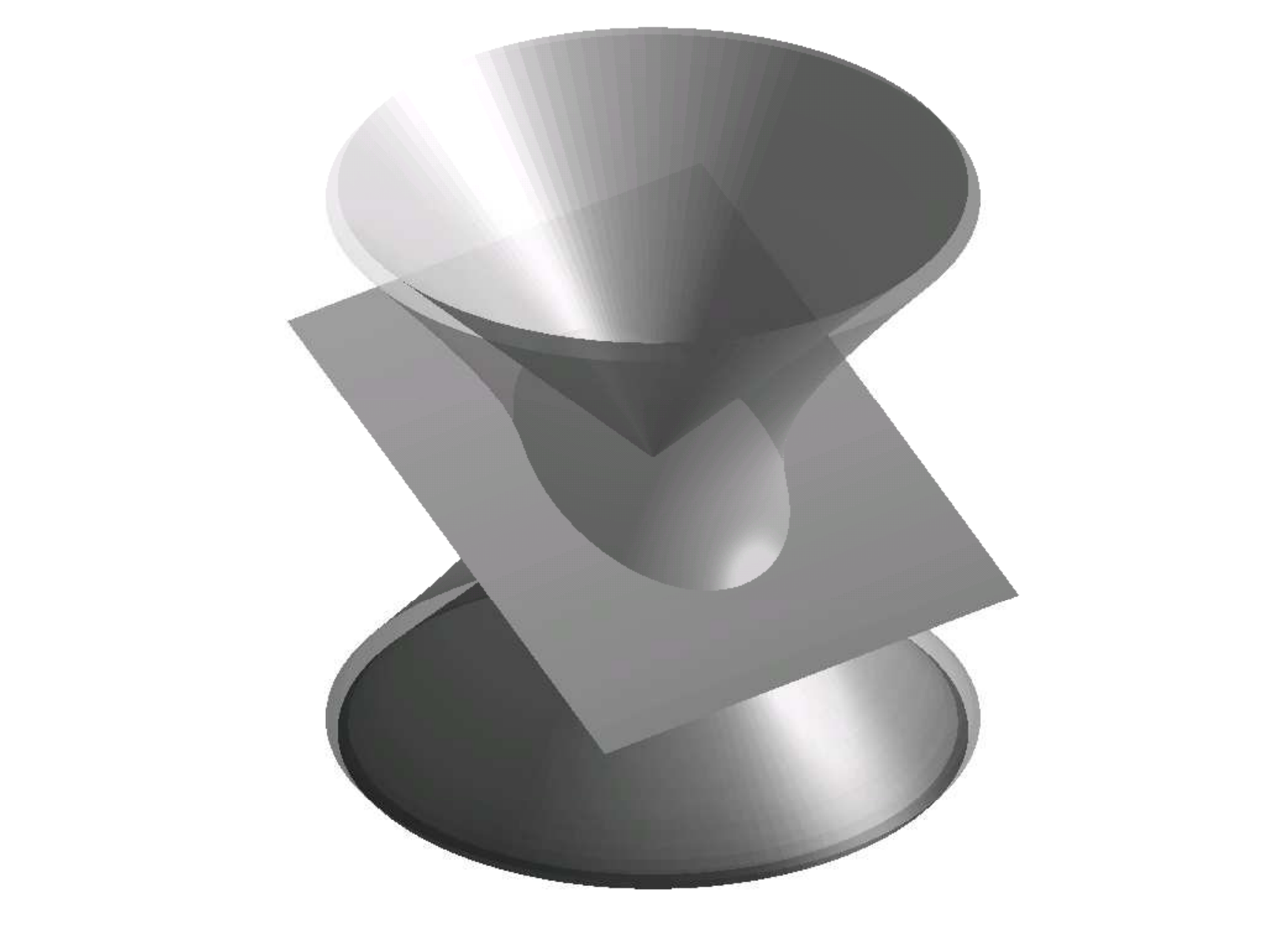}
\caption{\em Construction of the geodesics of the de Sitter
manifold.}
\end{center}
\end{figure}
\begin{eqnarray}
C^+ =\{X\in M(1,4) , \eta_{AB} X^A X^B =0, X^0>0 \}. \label{cone}
\end{eqnarray}
Each two-plane intersects $C^+$ either along a pair of generatrices, or along a single generatrix or only at the vertex. This give the
timelike, lightlike and spacelike geodesics, respectively. For later purposes, we exploit this property to provide intrinsic
characterizations of the timelike and lightlike geodesics. This will prove useful in the applications since timelike and lightlike
geodesics represent the worldlines of free massive and, respectively, massless particles. Given a timelike geodesic we
select two future oriented lightlike five-vectors $\xi$ and $\eta$ lying respectively on one and the other of the two pertinent
generatrices (see figure 1). Then, it is not too difficult to show that the geodesic can be parameterized as
\begin{eqnarray}
X(\tau) =R\frac {\xi e^{\frac {c\tau}R}-\eta e^{\frac
{-c\tau}R}}{\sqrt {2\xi\cdot \eta}},\label{geodesic}
\end{eqnarray}
where $\tau$ is the proper time. Denoting by $m$ the mass of the free particle whose worldline is assumed to be given by
(\ref{geodesic}), we normalize the dimensionless vectors $\xi$ and $\eta$ according to
\begin{eqnarray}
\xi\cdot\eta =\frac {2m^2}{k^2},\label{normalization}
\end{eqnarray}
where $k$ is some constant having the dimensions of a mass and whose value can be chosen time by time according to the specific
convenience. With this normalization, formula (\ref{geodesic}) reads
\begin{equation}
X(\tau) =\frac {kR}{2m} \left( \xi e^{\frac {c\tau}R}-\eta e^{\frac {-c\tau}R}  \right).\label{126}
\end{equation}
Once the constant $k$ has been fixed, we still have the freedom to change the separate normalization of $\xi$ and $\eta$ according to
the replacements $\xi \rightarrow \mu \xi$, $\eta \rightarrow \mu^{-1} \eta$ ($\mu>0$) which, as (\ref{126}) shows, corresponds to
a shift of the zero of the proper time. Using (\ref{normalization}) and introducing the initial point on the geodesic
\begin{eqnarray}
X_0=:X(0)=\frac {kR}{2m} (\xi-\eta), \label{initial}
\end{eqnarray}
we can give an alternative parametrization of it as
\begin{eqnarray}
X(\tau)=X_0 e^{-\frac {c\tau}R} +\frac {kR}{m} \xi \sinh \frac {c\tau}R .\label{128}
\end{eqnarray}
The normalization (\ref{normalization}) and equation (\ref{initial}) imply
\begin{equation}
\xi\cdot X_0 =-\frac {Rm}{k}.\label{129}
\end{equation}
For a massless particle, which moves along a lightlike geodesic, formulas (\ref{geodesic}) and (\ref{126}) loose their meaning since
$m=0$ and the vectors $\xi$ and $\eta$ degenerate into a single lightlike vector $\xi$ (see figure 1). However, we can use equation (\ref{128}) to
perform the limit of a lightlike geodesic. Since along a lightlike curve $d\tau =ds/c=0$, we set
\begin{eqnarray}
m=k\epsilon, \qquad \tau =\frac \sigma{c}\epsilon,
\end{eqnarray}
and let $\epsilon \rightarrow 0$ in eqns. (\ref{128}) and (\ref{129}) while keeping $k$ fixed. Then we find
\begin{eqnarray}
X(\sigma)=X_0+\sigma \xi, \qquad \xi \cdot X_0=0,\label{nullgeod}
\end{eqnarray}
where $\sigma$ is an affine parameter. This formula gives the equation of a lightlike geodesic as function of an initial point
$X_0$ and a lightlike vector $\xi$. Therefore, while timelike geodesics are certainly not straight lines in $M(1,4)$, a lightlike
geodesic is a straight line, lying on the Minkowski hyperboloid. It is characterized by one lightlike five-vector which is parallel to
the geodesic and by the choice of an initial event which uniquely selects the particular geodesic among the infinitely many pointing
in that direction.\\
\indent As pointed out in the introduction, due to the presence of curvature, no privileged class of reference frames exists
on the Minkowski hyperboloid and any choice of a local coordinate system is in principle acceptable for the description of
physical phenomena. However, it is still possible to select on $dS(1,3)$ certain classes of coordinate patches which have a high
degree of symmetry and which have a well defined cosmological significance. These are the so called flat, spherical and
hyperbolical coordinates, which we list below.

%%%%%%%%%%%%%%%%%%%%%%%%%%%%%%%%%%%%%%%%%%%%%%%%%%%%
\subsection{Flat coordinates}
The flat coordinates \cite{Lemaitre} $\{x^\mu; \mu=0,1,2,3; x^0=ct, x^i, i=1,2,3\}$ are defined by
\begin{eqnarray}
&& X^0 (x^\mu)= R\sinh \frac {ct}R +\frac {\vec x^2}{2R}\, e^{\frac {ct}R}\ ;  \cr
&& X^i (x^\mu)= e^{\frac {ct}R} x^i \ , \qquad i=1,2,3\ ; \label{flatland} \\
&& X^4 (x^\mu)= R\cosh \frac {ct}R -\frac {\vec x^2}{2R}\, e^{\frac
{ct}R}\ , \nonumber
\end{eqnarray}
where we have used the notation $\vec x=(x^1,x^2,x^3)$, and $\vec x^2=\delta_{ij} x^i x^j$.
They cover the half of the de Sitter manifold with $X^0+X^4>0$. The boundary of this region is the cosmological horizon of the (localized)
geodesic observer at rest at the origin, namely of the one whose geodesic is given by the equation
\begin{eqnarray}
Y(\tau)=R\frac {ue^{\frac {c\tau}R}-ve^{\frac {-c\tau}R}}{\sqrt{2u\cdot v}}, \label{ipsilon}
\end{eqnarray}
where $u=(1,0,0,0,1)$ and $v=(1,0,0,0,-1)$. In flat coordinates the metric (\ref{122}) takes the form
\begin{eqnarray}
ds^2 =g_{\mu\nu}dx^\mu dx^\nu=c^2 dt^2 -a^2(t) \delta_{ij} dx^i dx^j
= c^2 dt^2 -e^{2\frac {ct}R} \delta_{ij} dx^i dx^j \ .\label{flatmetric}
\end{eqnarray}
Fixed time $t=\bar t$ regions correspond to a $X^0+X^4=constant$ slice (see figure 2).

\begin{figure}[h]
\begin{center}
\includegraphics[height=6cm]{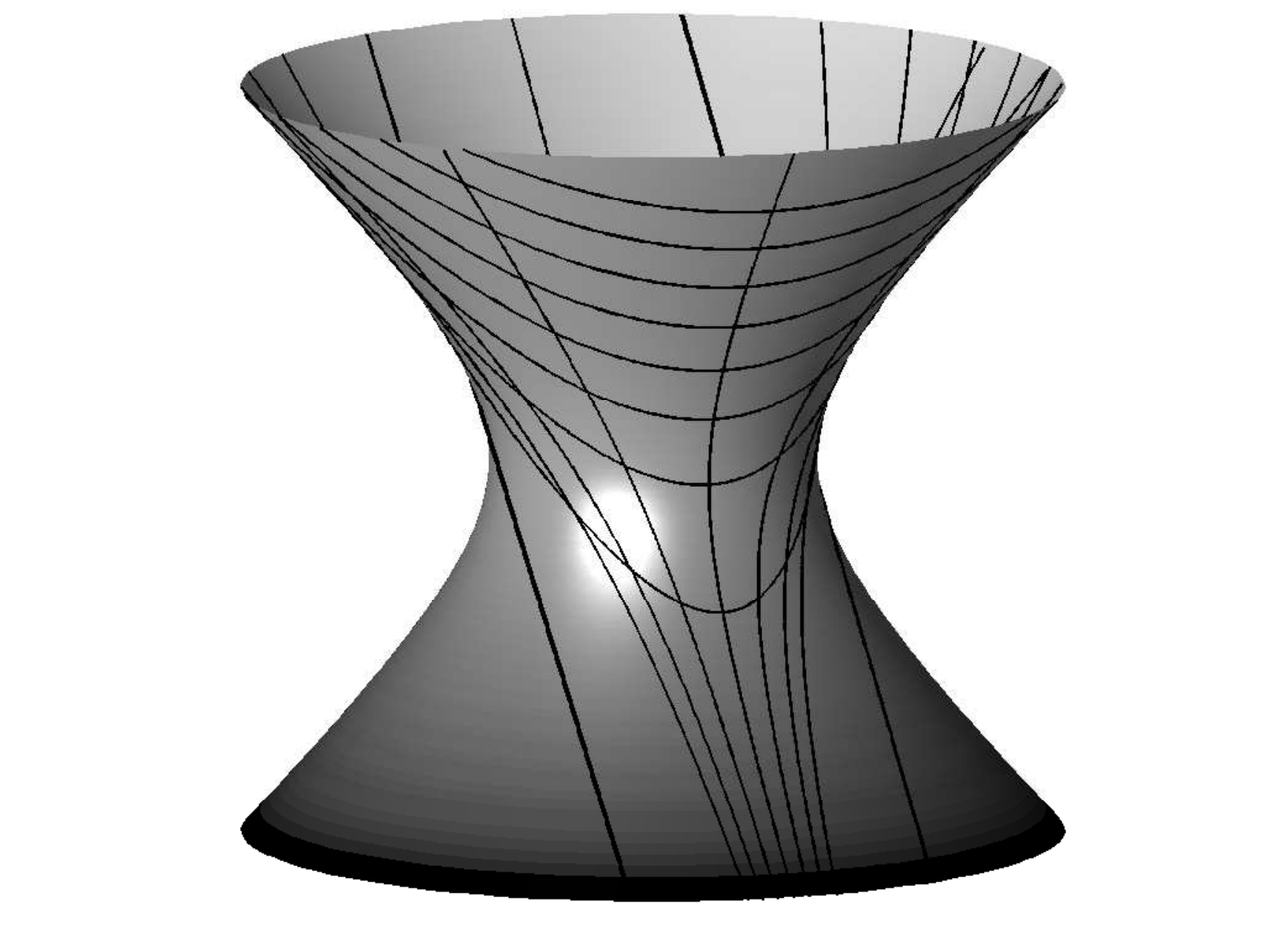}
\caption{\em Space and time sections for the flat coordinates}
\end{center}
\end{figure}

Such slices are endowed with the spatially flat Euclidean metric rescaled by a factor $e^{\frac {2ct}R}$. To an
observer using such coordinates de Sitter space appears as a universe consisting of flat spatial slices expanding exponentially in time. \\
Computing the curvature tensor as in appendix B, we find
\begin{eqnarray}
{\RR^{\mu\nu}}_{\rho\sigma}= \frac 1{R^2}
(\delta^\mu_\rho \delta^\nu_\sigma-\delta^\mu_\sigma
\delta^\nu_\rho) \ ,
\end{eqnarray}
which shows that $dS(1,3)$ is a space with constant negative Gaussian curvature $K=-1/R^2$. The Ricci tensor and the curvature
scalar are
\begin{eqnarray}
&& \RR_{\mu\nu} = \frac 3{R^2} g_{\mu\nu} \ , \quad\  \RR=\frac {12}{R^2} \ .
\end{eqnarray}

%%%%%%%%%%%%%%%%%%%%%%%%%%%%%%%%%%%%%%%%%%%%%%%%%%%%%%%%%%
\subsection{Spherical coordinates}
The spherical coordinates \cite{Lanczos} are defined as $\{x^\mu; \mu=0,1,2,3; x^0=ct, x^1=\chi,
x^2=\theta, x^3=\phi; 0<\chi<\pi, 0<\theta <\pi, 0<\phi <2\pi\}$
with
\begin{eqnarray}
&& X^0 (x^\mu)= R\sinh \frac {ct}R \ ; \quad\ X^1 (x^\mu)= R\cosh \frac {ct}R \sin \chi \sin \theta \sin \phi \ ; \cr
&& X^2 (x^\mu)= R\cosh \frac {ct}R \sin \chi \sin \theta \cos \phi \ ; \quad\  X^3 (x^\mu)= R\cosh \frac {ct}R \sin \chi \cos \theta \ ; \\
&& X^4 (x^\mu)= R\cosh \frac {ct}R \cos \chi \ .\nonumber
\end{eqnarray}
In spherical coordinates the metric takes the form
\begin{eqnarray}
ds^2 = c^2 dt^2 -R^2 \cosh^2 {\frac {ct}R} \left[d\chi^2 +\sin^2
\chi (d\theta^2+ \sin^2 \theta d\phi^2)\right] \ .
\end{eqnarray}
Fixed time $t=\bar t$ regions correspond to constant $X^0$ values $R\sinh \frac {ct}R$ and hence to three-dimensional spheres of radius
$R\cosh \frac {ct}R$ (see figure 3)
\begin{equation}
\sum_{a=1}^4 (X^a)^2=R^2 \cosh^2 {\frac {ct}R} \ .
\end{equation}

\begin{figure}[h]
\begin{center}
\includegraphics[height=6cm]{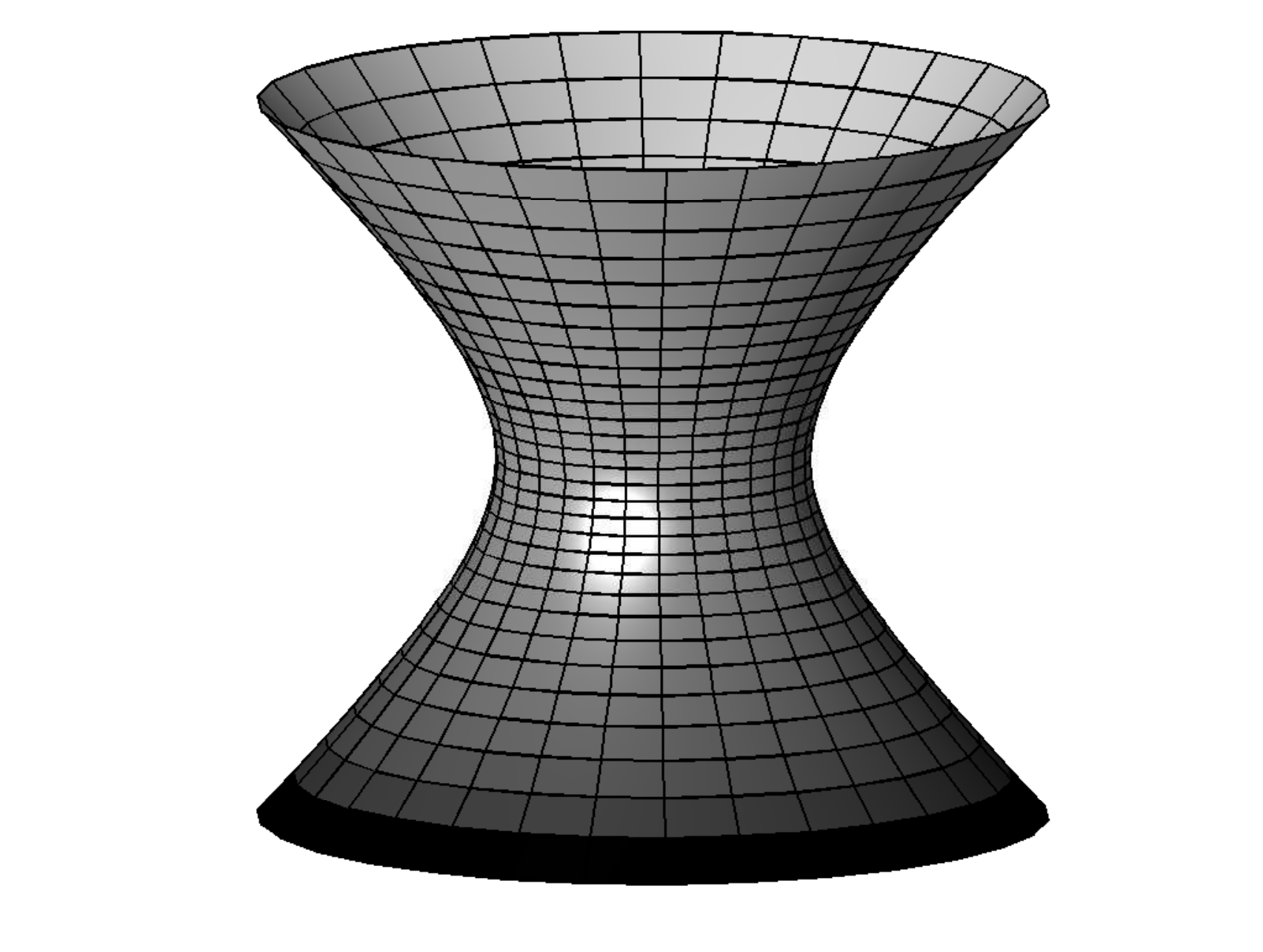}
\caption{\em Space and time sections for the spherical coordinates}
\end{center}
\end{figure}

Thus, to an observer using spherical coordinates, the Minkowski hyperboloid appears as a closed spherical universe which expands in
time as $\cosh \frac {ct}R$.

%%%%%%%%%%%%%%%%%%%%%%%%%%%%%%%%%%%%%%%%%%%%%%%%%%%%%%%%%%%%%%%%%%%%
\subsection{Hyperbolic coordinates}
The hyperbolic coordinates are defined as $\{x^\mu; \mu=0,1,2,3;
x^0=ct, x^1=\chi, x^2=\theta, x^3=\phi; 0<\chi, 0<\theta <\pi,
0<\phi <2\pi\}$ with
\begin{eqnarray}
&& X^0 (x^\mu)= R\sinh \frac {ct}R \cosh \chi \ ; \quad\ X^1 (x^\mu)= R\sinh \frac {ct}R \sinh \chi \sin \theta \sin \phi \ ; \cr
&& X^2 (x^\mu)= R\sinh \frac {ct}R \sinh \chi \sin \theta \cos \phi \ ; \quad\ X^3 (x^\mu)= R\sinh \frac {ct}R \sinh \chi \cos \theta \ ; \\
&& X^4 (x^\mu)= R\cosh \frac {ct}R \ ,\nonumber
\end{eqnarray}
and the metric reads
\begin{eqnarray}
ds^2 = c^2 dt^2 -R^2 \sinh^2 {\frac {ct}R} \left[d\chi^2 +\sinh^2
\chi (d\theta^2+ \sin^2 \theta d\phi^2)\right] \ .
\end{eqnarray}
Here fixed time regions correspond to threedimensional hyperbolic spatial sections (with constant curvature) which expand in time as
$\sinh \frac {ct}R$ (see figure 4).

\begin{figure}[h]
\begin{center}
\includegraphics[height=6cm]{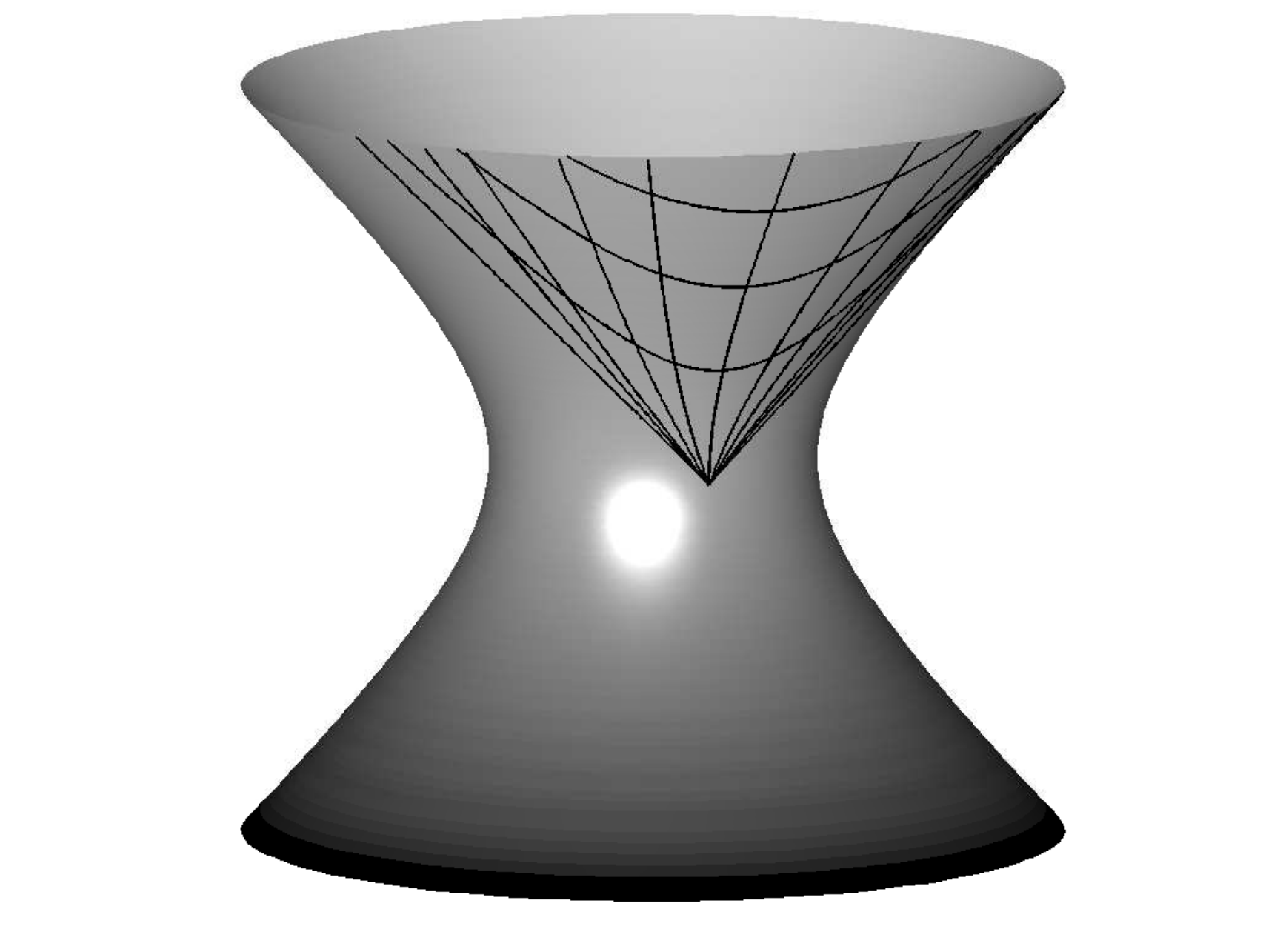}
\caption{\em Space and time sections for the hyperbolic coordinates}
\end{center}
\end{figure}

The three coordinate systems above conform to the cosmological principle and correspond to the well known flat, closed and open
Robertson-Walker metric respectively. In de Sitter space they are all on the same footing. However, since in our actual matter-energy
filled universe all observations are consistent with the total energy density being critical and therefore with physical space
being flat, we shall work in the following with the flat coordinates (\ref{flatland}).

%%%%%%%%%%%%%%%%%%%%%%%%%%%%%%%%
\section{Flat limit}\label{ssec:flatlimit}
We wish to show in which precise geometrical sense we can expect de Sitter space to go over to flat Minkowski space, and the de
Sitter algebra to contract (\`a la In\"on\"u-Wigner \cite{InonuWigner}) into the Poincar\'e algebra, when $R\rightarrow
\infty$, namely in the limit of zero curvature. In particular, we shall check that in this limit the flat coordinates
(\ref{flatland}) go over to the ordinary inertial ones in $M(1,3)$. Obviously, such a limit must be performed
in some precise manner, and it cannot be thought of in the usual sense of continuity. Indeed, under this limit the topology of the
manifold must change because the topologies of the hyperbolic and of the flat Minkowski spaces are not the same. Something similar
happens in the case of a two-dimensional Euclidean sphere of radius $R$, for which the isometry group is the ordinary rotation group
$SO(3)$ about the center of the sphere. If we wish to consider the limit when the radius of the sphere goes to infinity, we must first
fix a point $p$ on the sphere which we understand conventionally to be the ``origin'', for example the point of intersection of the
sphere with the $x^3$ axis. Then we shift the origin of the cartesian coordinate system from the center of the sphere to the
point $p$, so that the tangent plane to the sphere at $p$ is now the $(x^1,x^2)$ plane. If we now let $R$ increase indefinitely the
sphere will approximate more and more the $(x^1,x^2)$ plane, and it is in this precise sense that the sphere approaches a flat plane as
$R$ diverges. At the same time, whereas a rotation about the $x^3$ axis remains such in the limit, a rotation by an angle $\alpha/R$
(with $\alpha$ fixed) about a direction parallel to the $x^1$ (respectively, $x^2$) axis degenerates in the limit in a translation
by $-\alpha$ (respectively, by $\alpha$) along the $x^2$ (respectively, $x^1$) axis. In this way we see how the
three-dimensional rotation group contracts in the limit $R\rightarrow \infty$ into the group of rotations and translations in the plane.\\
\indent
In the case of interest to us things work as follows. In analogy to the procedure just outlined for the sphere, we shift the origin of
the inertial coordinates $X^A$ in $M(1,4)$ from the center of the Minkowski hyperboloid to the point $\bar X=(0,0,0,0,R)$ which we
have earlier chosen as the origin in $dS(1,3)$. Once this is done, equations (\ref{flatland}) can be rewritten as
\begin{eqnarray}
&& X^0 (x^\mu)= R\sinh \frac {x^0}R +\frac {\vec x^2}{2R}\, e^{\frac {x^0}R}\ ; \cr
&& X^i (x^\mu)= e^{\frac {x^0}R} x^i \ , \qquad i=1,2,3\ ; \label{reflat}\\
&& X^4 (x^\mu)= R\left(\cosh \frac {x^0}R-1\right) -\frac {\vec x^2}{2R}\, e^{\frac {x^0}R}\ . \nonumber
\end{eqnarray}
The hyperplane tangent to the hyperboloid at the point $\bar X$ is now the $(X^0, X^1, X^2, X^3)$ hyperplane, which we can identify
geometrically with the familiar flat Minkowski space $M(1,3)$. Then, if we let $R$ increase indefinitely, $dS(1,3)$ will approximate
$M(1,3)$ better and better, and will eventually identify with it in the limit $R\rightarrow \infty$. As to the corresponding behavior
of the transformations of the de Sitter group we distinguish several cases.
\begin{enumerate}
\item {Rotations in three space $(X^1, X^2, X^3$):}\\
This is $X^{\prime i}=\Omega^i_{\ \ j}\, X^j$, where $\Omega\in SO(3)$. Expressing $X^i$ in terms of the flat coordinates, this
simply becomes $x^{\prime i}=\Omega^i_{\ \ j}\, x^j$ and, in particular $\vec x^{\prime2}=\vec x^{2}$. Therefore, the effect of
such a rotation on (\ref{reflat}) is
\begin{eqnarray}
&& X^{\prime 0} =R \sinh \frac {x^{\prime 0}}R +\frac {\vec x^2}{2R}e^{\frac {x^{\prime0}}R} =X^0=R \sinh \frac {x^{0}}R +\frac
{\vec x^2}{2R}e^{\frac {x^{0}}R}, \cr
&& X^{\prime i} =e^{\frac {x^{\prime0}}R} x^{\prime i}=\Omega^i_{\ \ j}\, X^j=e^{\frac {x^0}R}\Omega^i_{\ \ j}\, x^j, \label{aaa} \\
&& X^{\prime4} =R\left( \cosh \frac {x^{\prime0}}R -1 \right) -\frac {\vec x^2}{2R} e^{\frac {x^{\prime0}}R} =X^4 =R\left( \cosh \frac
{x^{0}}R -1 \right) -\frac {\vec x^2}{2R} e^{\frac {x^{0}}R}. \nonumber
\end{eqnarray}
In the limit $R\rightarrow \infty$ it becomes
\begin{eqnarray}
&& x^{\prime0} =x^0 \ , \label{prima} \quad\ x^{\prime i}=\Omega^i_{\ \ j}\, x^j\ ,
\end{eqnarray}
with $X^{\prime 4}=X^4=0$, and this is a rotation in $M(1,3)$.
\item{Rotations in the $i$--$4$ plane:}\\
We consider a rotation by an angle $\lambda/R$, with $\lambda$ fixed:
\begin{eqnarray}
&& X^{\prime0}=X^0  \cr
&& X^{\prime i}=X^i \cos \frac \lambda{R} -X^4 \sin \frac \lambda{R}-R\sin \frac \lambda{R} \label{rotaz} \\
&& X^{\prime4} =X^i \sin \frac \lambda{R}+X^4 \cos \frac \lambda{R}-R\left(1-\cos \frac\lambda{R} \right). \nonumber
\end{eqnarray}
Taking  (\ref{reflat}) into account, in the limit $R\rightarrow\infty$ this becomes
\begin{eqnarray}
&& x^{\prime0}=x^0 \quad\  x^{\prime i}=x^i-\lambda,
\end{eqnarray}
which is a translation by $\lambda$ along the axis $x^i$ in $M(1,3)$.
\item{Boost along the direction $X^i$:}\\
This is given by
\begin{eqnarray}
&& X^{\prime0}=X^0 \cosh \beta+X^i \sinh \beta, \quad\ X^{\prime i}=X^0 \sinh \beta +X^i \cosh \beta,\label{ddd}
\end{eqnarray}
where $\beta$ is the boost parameter. We easily see from (\ref{reflat}) that in the limit
$R\rightarrow \infty$ this translates into the same boost in the $x^\mu$ coordinates:
\begin{eqnarray}
&& x^{\prime0}=x^0 \cosh \beta+x^i \sinh \beta, \quad\ x^{\prime i}=x^0 \sinh \beta +x^i \cosh \beta.
\end{eqnarray}
\item{Boost along the direction $X^4$:}\\
We consider a boost by a hyperbolic angle $\mu/R$ with $\mu$ fixed:
\begin{eqnarray}
&& X^{\prime0}=X^0 \cosh \frac \mu{R} +X^4\sinh \frac \mu{R} +R\sinh \frac \mu{R},\cr
&& X^{\prime i}=X^i, \label{eee} \\
&& X^{\prime4}=X^0\sinh \frac \mu{R} +X^4\cosh \frac \mu{R} +R\left(\cosh \frac \mu{R}-1 \right).\nonumber
\end{eqnarray}
Taking always into account (\ref{reflat}), in the limit $R\rightarrow \infty$ this becomes
\begin{eqnarray}
&& x^{\prime 0}=x^0+\mu \quad\ x^{\prime i}=x^i. \label{ultima}
\end{eqnarray}
Thus a boost along the $X^4$ direction becomes a translation in time.
\end{enumerate}
Collecting formulas (\ref{prima})--(\ref{ultima}) we find, as expected, that in the limit $R\rightarrow \infty$ the flat
coordinates on the Minkowski hyperboloid become inertial coordinates on $M(1,3)$ and the $SO(1,4)$ transformations become the Poincar\'e
ones. We have arrived at this result using the linear action of $SO(1,4)$ on the inertial coordinates $X^A$ on $M(1,4)$.
Alternatively, solving equations (\ref{flatland}) for the flat coordinates and obtaining
\begin{eqnarray}
&& x^0=R\log \left( \frac {X^0+X^4}R\right),\label{174}\quad\ x^i=e^{-\frac {x^0}R} X^i=\frac {RX^i}{X^0+X^4},
\end{eqnarray}
we can directly express the $SO(1,4)$ transformations as transformations on the coordinates $x^\mu$. There is no use in
writing them down since they are complicated and nonlinear and not particularly illuminating. However, their limit as
$R\rightarrow\infty$ can be easily computed to give the Poincar\'e group on $M(1,3)$.\\
\indent
We conclude this section by exhibiting explicitly the differential form (\ref{formule}) of the generators and their $R\rightarrow
\infty$ limit. Consider for example a rotation (\ref{rotaz}) in the $i$--$4$ plane by a small angle
$\delta \lambda/R$. Then, to first order in $\delta\lambda$ and using formula (\ref{transform}), we have
\begin{eqnarray}
&& f(X)-\frac {\delta\lambda}R (N^{i4} f) (X)=f\left(X^0,X^i-X^4\frac {\delta \lambda}R -\delta \lambda, X^4+X^i\frac {\delta\lambda}R \right)\cr
&& \qquad\qquad\qquad\ =f(X)-\delta\lambda \left[ \frac {X^i}R \frac \partial{\partial X_4}-\left(\frac
{X^4}R+1\right)\frac \partial{\partial X_i} \right] f(X),
\end{eqnarray}
so that
\begin{eqnarray}
P_i=\frac 1R N_{i4} =\frac 1R N^{i4} =\frac {X^i}R \frac \partial{\partial X_4} -\left(\frac {X^4}R +1 \right) \frac
\partial{\partial X_i}. \label{166}
\end{eqnarray}
Note that in formula (\ref{166}) there appears the extra term $\partial/\partial X^i$ which would not result were we to apply
formula (\ref{generators}). The obvious reason for this is that the rotation (\ref{rotaz}) contains a non homogeneous part
which comes from having shifted the origin of the $X^A$ coordinates from the center of the hyperboloid to the origin $\bar
X=(0,0,0,0,R)$. The presence of this extra term is essential for the correct computation of the limiting generator. Indeed, if this term
were not there, the operator $P_i$ would vanish in the limit $R\rightarrow \infty$. Instead, we correctly have
\begin{eqnarray}
P_i \ \ {}_{\overrightarrow{R\rightarrow\infty}} -\frac \partial{\partial X_i}=\frac \partial{\partial X^i},
\end{eqnarray}
which is the differential generator of a translation along the axis $X^i$ in $M(1,3)$. The remaining generators are calculated in a
similar manner from the infinitesimal forms of (\ref{aaa}), (\ref{ddd}), (\ref{eee}). Summarizing, we have
\begin{eqnarray}
&& \frac Hc \!=\!\frac 1R N_{04}\!=\! \frac {X^0}R \frac \partial{\partial X^4}\!+\!\left(\! \frac{X^4}R \!+\!1\! \right)\!
\frac \partial{\partial X^0}, \quad
P_i \!=\!\frac 1R N_{i4}\!=\! -\frac {X^i}R \frac \partial{\partial X^4}\!+\!\left(\! \frac{X^4}R \!+\!1 \!\right)\! \frac \partial{\partial X^i},\cr
&& cK_i\!=\!N_{0i}\!=\!X^0 \frac \partial{\partial X^i}\!+\!X^i \frac \partial{\partial X^0},\quad
J_i\!=\!\frac 12 \varepsilon_{ijk} N^{jk}\! =\!\frac 12 \varepsilon_{ijk}\! \left(\! X^k \frac \partial{\partial X^j}\!-\!X^j \frac \partial{\partial
X^k}\!\right). \label{acca}
\end{eqnarray}
The generators $K_i$ and $J_i$ are the same as those appearing in (\ref{formule}) and calculated using (\ref{generators}). On the
contrary, compared to those calculated from (\ref{generators}), the generators $H$ and $P_i$ contain the extra terms $-\partial/\partial
X^0$ and $\partial/\partial X^i$ which arise from the operation of shifting the origin of the $X^A$ coordinates to $\bar X$. One easily
checks that, in spite of the presence of these extra terms, the commutation relations (\ref{97})--(\ref{104}) are satisfied. In the
limit $R\rightarrow \infty$ the generators (\ref{acca}) of the de Sitter group go over into those of the Poincar\'e group
and, due to (\ref{reflat}), can be expressed in terms of the flat coordinates (the inertial ones in $M(1,3)$):
\begin{eqnarray}
&& H=\frac \partial{\partial t},\quad P_i=\frac \partial{\partial x^i},\quad K_i=t\frac \partial{\partial x^i}+\frac {x^i}{c^2} \frac
\partial{\partial t},\quad J_i=\frac 12 \varepsilon_{ijk} \left( x^k \frac \partial{\partial x^j} -x^j \frac \partial{\partial x^k} \right).
\end{eqnarray}
In turn, if we perform the limit $c\rightarrow \infty$ in the formulas above, we obtain the generators of the Galilei group.
However, note that the limits $R\rightarrow\infty$ and $c\rightarrow\infty$ cannot be interchanged in
(\ref{97})--(\ref{104}). A different parametrization of the Lie algebra of $SO(1,4)$ is needed in order to obtain the commutation
relations of the Lie algebra of the Newton--Hooke group, which describes non relativistic mechanics in three-dimensional space with
positive constant curvature \cite{Bacry}.

%%%%%%%%%%%%%%%%%%%%%%%%%%%%%%%%%%%%%%%%%%%%%%%%%%%%%%%%%%%%%%%%%%%%%%%%%%%%%%%%%%%%%%%%%%%%%%%%%%%%%%%%%
\section{Kinematics in the Minkowski hyperboloid}\label{sec:kinematics}
The lack of a privileged set of coordinate systems in de Sitter space calls for the attempt to express physical
quantities and physical laws, whenever possible, in an intrinsic way, avoiding making reference to any particular coordinate patch.
In the preceding section we have already applied this program to the characterization of geodesic curves. This is important, since
timelike and lightlike geodesics describe the motion of free massive and massless classical pointlike particles, respectively (see
appendix B). In this section we elaborate further on this concept by first characterizing the geodesic motion of free particles in terms
of an intrinsic formulation of the associated conservation laws. In particular, we apply this notion to the definition of a conserved
energy and momentum of a de Sitter particle. In the second part of the section we study the scattering and decay processes of classical
pointlike particles in terms of the conservation of the total sum of the particle invariants at the collision point. With minor modifications,
the results of this section form essentially the content of a recent paper by us \cite{GKCM}.\\
\indent
The geodesic equation for a massive particle can be derived imposing the stationarity of the free particle action:
\begin{eqnarray}
S(\gamma)=-mc \int_\gamma ds =-mc\int_\gamma \sqrt {\eta_{AB} dX^A dX^B|_{X^2=-R^2}}.\label{freeparticle}
\end{eqnarray}
Since the de Sitter group is the group of isometries of the Minkowski hyperboloid, the action is invariant under the
transformations of the group. Then, given an infinitesimal transformation
\begin{eqnarray}
X^C \rightarrow X^C+\varepsilon (M^{AB})^C_{\ \ D} X^D,
\end{eqnarray}
where the matrices $M^{AB}$ are given in (\ref{generators}), Noether's theorem gives the corresponding conserved quantities
\begin{eqnarray}
K_{AB} =\frac mR (X_A(\tau) W_B(\tau)-X_B(\tau) W_A(\tau))=\frac 1R (X_A(\tau) \Pi_B(\tau)-X_B(\tau) \Pi_A(\tau)), \label{kappa}
\end{eqnarray}
where $\tau \rightarrow X^A(\tau)\in dS(1,3)$ is the particle geodesic curve (\ref{geodesic}) parameterized by the proper time
$d\tau=ds/c$, $W^A(\tau)=dX^A(\tau)/d\tau$ is the particle five-velocity and $\Pi^A(\tau)$ is the particle five-momentum. Since
\begin{eqnarray}
X^2(\tau)=\eta_{AB} X^A(\tau) X^B(\tau)=-R^2,
\end{eqnarray}
we have
\begin{eqnarray}
X(\tau)\cdot W(\tau)=\eta_{AB} X^A(\tau) W^B(\tau)=0,
\end{eqnarray}
namely the five-velocity, which is at each instant tangent to the curve $X^A(\tau)$, is also orthogonal to the five-position. From
equation (\ref{geodesic}) we get
\begin{eqnarray}
W(\tau) =c\frac {\xi e^{\frac {c\tau}R}+\eta e^{-\frac {c\tau}R}}{\sqrt {2\xi \cdot \eta}}=\frac {k c}{2m} (\xi e^{\frac
{c\tau}R}+\eta e^{-\frac {c\tau}R})\label{wditau}
\end{eqnarray}
and, solving for $\xi$ and $\eta$,
\begin{eqnarray}
&& \xi e^{\frac {c\tau}R}=\frac mk \!\! \left[\! \frac 1c W(\tau)+\frac 1R X(\tau) \right]=\xi(\tau), \label{eta} \quad
\eta e^{-\frac {c\tau}R}=\frac mk \!\! \left[\! \frac 1c W(\tau)-\frac 1R X(\tau) \right]=\eta(\tau).
\end{eqnarray}
Inserting (\ref{geodesic}) and (\ref{wditau}) into (\ref{kappa}) gives
\begin{eqnarray}
K_{AB}=mc \frac {\xi_A \eta_B -\xi_B \eta_A}{\xi\cdot \eta}.\label{195}
\end{eqnarray}
These are the components of the two-form
\begin{eqnarray}
K_{(\xi,\eta)} =mc\frac {\xi\wedge \eta}{\xi\cdot \eta}
\end{eqnarray}
relative to the basis $(e_A)^B=\delta_A^{ B}$ in the ambient space $M(1,4)$ ($X=X^A e_A$). With the choice of the normalization
(\ref{normalization}) for the lightlike five-vectors $\xi$ and $\eta$ we have
\begin{eqnarray}
K_{AB}=\frac {k^2 c}{2m}(\xi_A \eta_B-\xi_B\eta_A),\quad K_{(\xi,\eta)} =\frac {k^2 c}{2m} (\xi\wedge \eta).\label{197}
\end{eqnarray}
Formulas (\ref{195})-(\ref{197}) provide an intrinsic characterization of the ten conserved quantities $K_{AB}$, expressed
in terms of the lightlike vectors $\xi$ and $\eta$ which uniquely identify the given geodesic. Of course, in the form (\ref{kappa}),
the dynamical variables $K_{AB}$ can be defined for any timelike curve $\tau\rightarrow X(\tau)$ in $dS(1,3)$, not necessarily a
geodesic, but then they will not be conserved in general. Inserting (\ref{initial}) in (\ref{197}) gives
\begin{eqnarray}
&& K_{AB}=\frac {k c}{R} \left( X_A(0) \xi_B-X_B (0) \xi_A \right)=\frac {k c}R e^{\frac {c\tau}R} \left( X_A(\tau)
\xi_B-X_B(\tau) \xi_A \right),\nonumber\\
&& K=K_{(\xi,X)} =\frac {k c}R e^{\frac {c\tau}R} \left( X(\tau) \wedge \xi \right),\label{198}
\end{eqnarray}
which is a second characterization of the conserved quantities, this time in terms of the lightlike vector $\xi$  and the initial point
$X(\tau)$ on the geodesic at any given initial proper time $\tau$ (for example $\tau=0$). The ten constants of the motion $K_{AB}$ are
not all independent. Indeed, once the vectors $\xi$ and $\eta$ have been separately normalized according to (\ref{normalization}) and
to the choice of the initial point $X(0)$ on the geodesic, they each have three independent components, so that $K_{AB}$ are specified by
the assignment of six independent constants. Relations (\ref{eta}) show that, as expected, these constants can, for
instance, be chosen as the initial position and the initial velocity of the particle. For a massless particle formula (\ref{195}) looses
its meaning, but (\ref{198}) retains its validity:
\begin{eqnarray}
K_{AB} =\frac {k c}R \left( X_A(0) \xi_B -X_B(0) \xi_A \right)=\frac 1R \left( X_A \Pi_B-X_B \Pi_A \right),\label{199}
\end{eqnarray}
where $\Pi^A=k c dX^A/d\sigma=k c \xi $ is the five-momentum of the particle.

%%%%%%%%%%%%%%%%%%%%%%%%%%%%%%%%%%%%%%%%%%%%%%%%%%%%%%%%%%%%%%%%%%%%%%%%%%
\subsection{Energy}\label{ssec:energy}
The action of the two-form $\xi\wedge \eta$ on a pair of five-vectors $a$ and $b$ is given by
\begin{eqnarray}
(\xi\wedge\eta)(a,b)=(\xi_A\eta_B-\xi_B\eta_A)a^Ab^B.
\end{eqnarray}
In particular, if we introduce in $M(1,4)$ the canonical orthonormal basis
$(e_A)^B=\delta_A^{B}$, we have
\begin{eqnarray}
(\xi\wedge\eta)(e_A,e_B)=\xi_A\eta_B-\xi_B\eta_A.
\end{eqnarray}
We are now ready to define the energy of a free particle whose geodesic motion is given by (\ref{geodesic}). Precisely, the energy
$E$ is by definition the conserved quantity associated to the invariance of the particle action (\ref{freeparticle}) under time
translation. Then, since the generator of time translation is given by $H=(c/R)N_{04}$ (see (\ref{formule})), we have
\begin{eqnarray}
E=cK_{04}=cK_{(\xi,\eta)} (e_0,e_4)=\frac {mc}R \left(X^4 \frac {dX^0}{d\tau}-X^0 \frac {dX^4}{d\tau}\right).\label{203}
\end{eqnarray}
This is the energy of the particle relative to the (localized) geodesic observer at rest at the origin, whose geodesic is given by equation
(\ref{ipsilon}). Then, since $u=e_0+e_4$ and $v=e_0-e_4$, we have
\begin{eqnarray}
E=:E_{(\xi,\eta)} (u,v) =-\frac {cK_{(\xi,\eta)}(u,v)}{u\cdot v}.\label{205}
\end{eqnarray}
Formula (\ref{205}) allows us to give an intrinsic (though obviously relative) general definition of the energy of a free de Sitter
particle. Contrary to the case of Einstein's relativity, where the energy of a particle is defined relative to a global inertial frame
(an inertial observer), in de Sitter relativity the lack of privileged global frames forces us to define the energy relative to
a strictly localized geodesic observer, and this is done as follows. We select as our localized geodesic observer a reference massive
free particle understood conventionally to be at rest, and denote by $\zeta$ and $\chi$ the future oriented lightlike vectors which
identify its timelike geodesic. Then, the energy of a de Sitter particle relative to such reference geodesic (the geodesic of the
particle ``at rest'') is defined as
\begin{eqnarray}
E=E_{(\xi,\eta)}(\zeta,\chi)=-\frac {c K(\zeta,\chi)}{\zeta\cdot\chi}.\label{206}
\end{eqnarray}
In particular, when the reference particle is the one localized at the origin we get back (\ref{205}). We note that
\begin{eqnarray}
E_{(\xi,\eta)}(\zeta,\chi)=E_{(\zeta,\chi)} (\xi,\eta) \label{207}
\end{eqnarray}
which expresses the symmetry between the active and the passive points of view: we can interpret (\ref{207}) either as the energy of
different $(\xi,\eta)$ particles relative to a fixed observer $(\zeta,\chi)$ (left hand side, active view) or as the energy of the
same particle $(\zeta,\chi)$ relative to different observers $(\xi,\eta)$ (right hand side, passive view). As expected, the
proper energy is $E_{(\xi,\eta)}(\xi,\eta)=mc^2$.\\
Using (\ref{initial}) and (\ref{198}) we can reexpress (\ref{205}) as
\begin{eqnarray}
E=\frac {kc^2}{R^2} (X_0\wedge \xi)(u,Y_0)=\frac {kc^2}{R^2} \left[ (u\cdot X_0)(\xi\cdot Y_0)-(X_0\cdot Y_0)(\xi\cdot u) \right],
\end{eqnarray}
where $X_0$ and $Y_0=(0,0,0,0,R)=\bar X$ are the initial points on the geodesics of the particle and, respectively, of the reference
particle. This formula holds true also when the particle is massless.\\
\indent
To our knowledge, definition (\ref{206}) traces back to G\"ursey \cite{gursey} where, however, it was expressed in stereographic coordinates.
Here we stress that, in the same way that there is a unique Lorentzian (i.e. relativistic) generalization of the kinetic energy
of a Galilean (i.e. nonrelativistic) particle, the de Sitter energy (\ref{206}) is the unique generalization of the Lorentzian energy.
From its expression (\ref{209}) in flat coordinates (see next subsection) we see that, contrary to the Galilean and the Lorentzian ones,
the de Sitter energy is not bounded below. However, this does not create problems, since it was proved in \cite{Abbott-Deser} in a
field--theoretical context, that the de Sitter energy is stable under a large class of perturbations.

%%%%%%%%%%%%%%%%%%%%%%%%%%%%%%%%%%%%%%%%%%%%%%%%%%%%%%%%%%%%%%%%%%%%%
\subsection{Energy in flat coordinates}\label{ssec:flatcoord}
If we introduce in $dS(1,3)$ some coordinate patch $(t,x^i)$, we write $X^A=X^A(t,x^i)$ and formula (\ref{203}) allows us to express
$E$ in terms of these local coordinates. For example, using (\ref{flatland}) we obtain the expression of
the energy of a massive particle in terms of the flat coordinates:
\begin{eqnarray}
E=mc\left(c+\frac {a^2(t)}R \vec x \cdot \vec v  \right)\frac {dt}{d\tau}=\frac {mc^2}{\sqrt {1-a^2(t)\frac {v^2}{c^2}}}-\frac cR
x^i p_i,\label{209}
\end{eqnarray}
where we have set $v^i=dx^i/dt$ and
\begin{eqnarray}
p_i =-\frac {ma^2(t) v^i}{\sqrt {1-a^2(t) \frac {v^2}{c^2}}}.\label{210}
\end{eqnarray}
In flat coordinates the particle action is given by
\begin{eqnarray}
S(\gamma)=-mc^2\int_\gamma dt \sqrt {1-e^{2\frac {ct}R} \frac {v^i v^j}{c^2} \delta_{ij}}. \label{211}
\end{eqnarray}
Taking into account that in flat coordinates the spatial distances dilate in the course of time by the exponential factor $e^{ct/R}$,
we have that an infinitesimal time translation is expressed by
\begin{eqnarray}
t\longrightarrow t-\varepsilon, \qquad\ x^i\longrightarrow x^i+\frac cR x^i \varepsilon.\label{212}
\end{eqnarray}
The action (\ref{211}) is invariant under the transformation (\ref{212}) and, by Noether's theorem, the corresponding conserved
quantity is precisely (\ref{209}). This result is of course a trivial consequence of the fact that an infinitesimal time
translation in $M(1,4)$, $X\rightarrow X+(c/R) \varepsilon M^{04} X$ (an infinitesimal boost along the direction $X^4$),
\begin{eqnarray}
&& X^0\longrightarrow X^0+\frac cR \varepsilon X^4 ,\quad\ X^i\longrightarrow X^i, \quad\ X^4\longrightarrow\frac cR \varepsilon X^0+X^4,
\end{eqnarray}
has precisely the expression (\ref{212}) in flat coordinates. Similarly, given the functions $X^A (t,x^i)$, using (\ref{203}), we
can find the explicit expression $E(t,x^i,dt/d\tau,dx^i/d\tau)$ of the energy in any coordinate system we like in $dS(1,3)$. Then we
express the action (\ref{freeparticle}) in terms of the coordinates $(t,x^i)$ and repeat in this coordinate system the above argument
about $E(t,x^i,dt/d\tau,dx^i/d\tau)$ being the conserved quantity associated to the invariance of the action under time translation.\\
For a massless particle formula (\ref{199}) gives
\begin{eqnarray}
&E&\!\!\!=cK_{04}=\frac {kc^2}R\left( X^4 \frac {dX^0}{dt}-X^0\frac {dX^4}{dt} \right)\frac {dt}{d\sigma}
=kc^2 \left( c+\frac {a^2(t)}R \vec x \cdot \vec v \right) \frac {dt}{d\sigma}\cr
&& \!\!\!=kc^3 \frac {dt}{d\sigma} -\frac cR x^i p_i, \qquad\ \ \  \mbox{    where    } \qquad\ \ \
p_i=-kce^{2\frac {ct}R} v^i \frac {dt}{d\sigma},
\end{eqnarray}
so we must find $dt/d\sigma$. From the relation $X^0+X^4=Re^{\frac {ct}R}$ and from (\ref{nullgeod}), we get
$\xi^0+\xi^4=dX^0/d\sigma+dX^4/d\sigma=ce^{\frac {ct}R}dt/d\sigma$ so that
\begin{eqnarray}
\frac {dt}{d\sigma} =\frac 1c e^{-\frac {ct}R} (\xi^0+\xi^4).
\end{eqnarray}
Therefore,
\begin{eqnarray}
E=kc^2 (\xi^0+\xi^4)e^{-\frac {ct}R} \left( 1 +\frac {e^{2\frac {ct}R}}{cR} \vec x \cdot \vec v \right)
\qquad\ \mbox{ and } \qquad\
p_i=-k(\xi^0+\xi^4)e^{\frac {ct}R} v^i.
\end{eqnarray}
In the flat limit $R\rightarrow\infty$ we have $E\rightarrow kc^2(\xi^0+\xi^4)$. Hence, if we associate a frequency
to the massless particle we get
\begin{eqnarray}
kc^2 (\xi^0+\xi^4)=h\nu,
\end{eqnarray}
and the final expressions for $E$ and $p_i$ are
\begin{eqnarray}
&& E=h\nu e^{-\frac {ct}R} \left( 1+\frac {e^{2\frac {ct}R}}{cR} \vec x \cdot \vec v \right),\label{220} \\
&& p_i=-\frac {h\nu}{c^2} e^{\frac {ct}R} v^i.\label{221}
\end{eqnarray}
In the flat limit formulas (\ref{209}) and (\ref{220}) go over into the usual expressions of the Minkowskian
energies of a massive and massless particle, respectively,
\begin{eqnarray}
E=\frac {mc^2}{\sqrt {1-\frac {v^2}{c^2}}}
\qquad\ \mbox{ and } \qquad\
E=h\nu.
\end{eqnarray}
In the next subsection we show, in particular, that formulas (\ref{210}) and (\ref{221}) represent, in the massive and
massless case, the covariant components in flat coordinates of the momentum of the particle.

%%%%%%%%%%%%%%%%%%%%%%%%%%%%%%%%%%%%%%%%%%%%%%%%%%%%%%
\subsection{Momentum}\label{ssec:momentum}
The momentum of a particle is the constant of motion associated to the invariance of the action under space translations.
This statement requires first fixing an origin in spacetime since, due to the presence of curvature, changing the origin affects
the definition of space translations. In addition, it presupposes that the spacetime manifold has been foliated into three-dimensional
spatial slices (parameterized by a suitable time coordinate) and that well defined space coordinates $x^i$ have been introduced
into each slice, defining three independent directions. In other words, spatial momentum can be defined componentwise only with
reference to some suitable coordinate system $(t,x^i)$ (with the convention that $(0,0)$ is the origin). Then, since we have decided
to work with flat coordinates, we shall define momentum relative to these particular coordinates. The origin $t=0$, $x^i=0$
is the point $\bar X$. The spatial slices of the flat coordinates are the hyperplanes of equation $X^0+X^4=Re^{\frac {ct}R}$
for each fixed value of $t$, and on any such slice the $x^i$ coordinates are proportional to the $X^i$. Then, the covariant momentum
components will be given by $p_i=\alpha K(v,e_i)$ $(i=1,2,3)$ where $\alpha$ is a positive normalization factor which must be chosen
in such a way that in the flat limit this expression reduces to the Minkowskian one. For this to hold we must
take $\alpha=1$. Thus,
\begin{eqnarray}
p_i=K(v,e_i)=K(e_0,e_i)-K(e_4,e_i)=K_{0i}-K_{4i}.\label{224}
\end{eqnarray}
Then, using (\ref{kappa}) and (\ref{199}), we have
\begin{eqnarray}
p_i=\frac mR \left( X^i \frac {dX^0}{dt}-X^0 \frac {dX^i}{dt}+X^i \frac {dX^4}{dt}-X^4 \frac {dX^i}{dt} \right)\frac {dt}{d\tau}
\end{eqnarray}
in the massive case and
\begin{eqnarray}
p_i=\frac {kc}R \left( X^i \frac {dX^0}{dt}-X^0 \frac {dX^i}{dt}+X^i \frac {dX^4}{dt}-X^4 \frac {dX^i}{dt} \right)\frac {dt}{d\sigma}
\end{eqnarray}
in the massless one. Using (\ref{flatland}) we finally find, in the two respective cases, expressions
(\ref{210}) and (\ref{221}). The corresponding contravariant components of the momentum are, respectively,
\begin{eqnarray}
p^i=\frac {mv^i}{\sqrt {1-a^2(t)\frac {v^2}{c^2}}}
\qquad \mbox{ and } \qquad\
p^i=\frac {h\nu}{c^2} e^{-\frac {ct}R} v^i.
\end{eqnarray}
In the flat limit $R\rightarrow\infty$ these expressions go over into the corresponding Minkowskian ones:
\begin{eqnarray}
p^i=\frac {mv^i}{\sqrt {1-\frac {v^2}{c^2}}},\qquad\ p^i=\frac {h\nu}c n^i.
\end{eqnarray}
As expected, expression (\ref{210}) yields the conserved quantities associated to the invariance of the action (\ref{211}) under
space translations $x^i\rightarrow x^i+\varepsilon^i$.
Here it is worth nothing that the generators of space translations in flat coordinates are the usual differential
operators $\partial/\partial x^i$ $(i=1,2,3)$ which commute among each other. This can also be seen directly from formula (\ref{224})
which tells us that, relative to flat coordinates, the generators of space translations are the operators
\begin{eqnarray}
\label{ticoni} T_i=\frac 1R (N_{i4}+N_{0i})=P_i+\frac cR K_i,
\end{eqnarray}
where $P_i$ and $K_i$ are given in (\ref{formule}). Indeed, from (\ref{generators}) and the definition (\ref{flatland})
and (\ref{174}) of flat coordinates we have
\begin{eqnarray}
&& T_i =\frac 1R \left( X_i \frac {\partial}{\partial X^4} -X_4 \frac {\partial}{\partial X^i}
+X_0 \frac {\partial}{\partial X^i}-X_i \frac {\partial}{\partial X^0} \right)
=\frac {\partial}{\partial x^i}. \label{qformulaq}
\end{eqnarray}
Likewise, the generator of time translations, expressed in flat coordinates, is given by
\begin{eqnarray}
\label{tizero} T_0=\frac cR N_{04}=\frac cR \left( X^0 \frac {\partial}{\partial X^4} +X^4 \frac {\partial}{\partial X^0} \right)
=\frac {\partial}{\partial t}-\frac cR x^i \frac {\partial }{\partial x^i}
\end{eqnarray}
and we have the following commutation relations
\begin{eqnarray}
[T_i, T_j]=0, \qquad [T_0,T_j]=\frac cR T_j \qquad (i,j=1,2,3).
\end{eqnarray}
Instead of using (\ref{224}) we could have employed the generators $P_i$ in (\ref{formule}) to define the momentum as
$p_i =K(e_i,e_4)=K_{i4}$. This corresponds to spatial sections given by $X^0=const.$ and therefore gives the covariant
components of the momentum in spherical coordinates.\\
Formulas (\ref{205}), (\ref{206}) and (\ref{224}) for energy and momentum of a particle are special cases of the general
expression $K_{(\xi,\eta)} (n,m)$ which, given two arbitrary non parallel five-vectors $n$ and $m$, furnishes the conserved
quantity associated to the invariance of the particle action under the infinitesimal de Sitter transformation in the $(n,m)$
plane generated by the twoform $n\wedge m$.

%%%%%%%%%%%%%%%%%%%%%%%%%%%%%%%%%%%%%%%%%%%%%%%%%%%%%%%%%%%%%%%%
\subsection{de Sitter mass shell}\label{ssec:massshell}
In Einstein's relativity energy and momentum are related by the well known formula
\begin{eqnarray}
E^2-\vec p^{\, 2} c^2=m^2c^4\label{230}
\end{eqnarray}
which defines the mass shell in momentum space (or equivalently, the Casimir operator of the Poincar\'e group involving
the mass). The corresponding formula in de Sitter relativity involves all conserved quantities and is readily obtained
by squaring (\ref{195}):
\begin{eqnarray}
K_{AB} K^{AB}=-2m^2 c^2.\label{231}
\end{eqnarray}
We expect of course (\ref{231}) to go over into (\ref{230}) in the degenerate flat limit. In order to check this it is convenient
to isolate the contributions of energy and momentum from the rest in (\ref{231}). This decomposition will depend on the choice
of the coordinate system. If we carry it out in the flat coordinates, we find that (\ref{231}) can be rewritten as
\begin{eqnarray}
E^2 -\vec p^{\, 2} c^2 +2\frac {c^2}R \sum_{i=1}^3 p_i L_{0i} -\frac {c^2}{2R^2} \sum_{i,j=1}^3 L_{ij} L^{ij}=m^2 c^4,
\end{eqnarray}
where
\begin{eqnarray}
&& L_{ij}=m \frac {dt}{d\tau} \left( X^i \frac {dX^j}{dt}-X^j \frac {dX^i}{dt} \right)=e^{\frac {ct}R} (x^i p^j -x^j p^i)=R K_{ij}\label{233}
\end{eqnarray}
and
\begin{eqnarray}
&&L_{0i}=\frac {mc}{\sqrt {1-a^2(t) \frac {v^2}{c^2}}}\left\{ (x^i-v^i t)+\left[ x^i \frac {\vec x \cdot \vec v}{Rc} e^{2 \frac {ct}R}
-v^i \frac {\vec x^2}{2Rc} e^{2\frac {ct}R}\right. \right. \cr
&& \qquad\ \left. \left. -\frac {v^i}c e^{\frac {ct}R} R \left( \sinh \frac {ct}R-e^{-\frac {ct}R} \frac {ct}R \right)
\right]\right\}=RK_{0i}. \label{234}
\end{eqnarray}
The term (\ref{233}) represents the angular momentum of the particle, whereas (\ref{234}) is the conserved quantity associated to
a boost (velocity of the center of mass). As $R\rightarrow \infty$, the two terms approach the finite limits $x^i p^j-x^j p^i$ and
$\frac {mc}{\sqrt {1-v^2/c^2}}(x^i -v^i t)$, respectively, and we recover the Minkowskian expression (\ref{230}).
%%%%%%%%%%%%%%%%%%%%%%%%%%%%%%%%%%%%%%%%%%%%%%%%%%%%%%%%%%
%%%%%%%%%%%%%%%%%%%%%%%%%%%%%%%%%%%%%%%%%%%%%%%%%%%%%%%%%%
\subsection{Collisions and decays}\label{collision&decay}
So far we have been concerned with the de Sitter kinematics of a single free classical point particle.
We now consider a general process
\begin{eqnarray}
b_1+b_2\longrightarrow c_1+c_2+\ldots+c_N \label{235}
\end{eqnarray}
in which two incoming particles of masses $m_i$ ($i=1,2$) collide at some given spacetime point $X_0$ giving rise to a certain
number of outgoing particles of masses $\tilde m_f$ ($f=1,2,\ldots,N$). Obviously, without the knowledge of the underlying interaction,
no information can be gained about the differential cross section of the various channels of the process. However,
due to spacetime homogeneity, at the collision point the total energy-momentum four-vector should be conserved:
\begin{eqnarray}
\pi_1^\mu+\pi_2^\mu =\sum_{f=1}^N \tilde \pi_f^\mu,\label{236}
\end{eqnarray}
where $\pi^\mu=mdx^\mu/d\tau$ (or $\pi^\mu =kc dx^\mu/d\sigma$) denotes the one particle energy-momentum relative to some local
coordinate system. Now, during the geodesic motion of the particle, $\pi^\mu$ is not conserved since it changes according to the
equation of parallel transport (see appendix B). However, we can easily recast (\ref{236}) into an equation involving only
the conserved quantities $K_{AB}$ of each particle. This has the advantage over (\ref{236}) of being expressed in intrinsic form.
To see this, note that for each particle, at the collision point $X_0$, we have
\begin{eqnarray}
K_{AB}=\left. \frac 1R \left( X_A \frac {\partial X_B}{\partial x^\mu} - X_B \frac {\partial X_A}{\partial x^\mu}\right)\right|_{x=x_0}
\pi^\mu (x_0), \qquad\ X_0=X(x_0). \label{237}
\end{eqnarray}
Then, by summing over the initial and final particles, it follows from (\ref{236}) that
\begin{eqnarray}
K_1+K_2=\sum_{f=1}^N \tilde K_f. \label{238}
\end{eqnarray}
A similar formula holds in the case of particle decay
\begin{eqnarray}
K=\sum_{f=1}^N \tilde K_f. \label{239}
\end{eqnarray}
Equation (\ref{238}) is a consequence of (\ref{236}), but it is in fact equivalent to it. To see this we assume the collision point $X_0$
to correspond to the common zero of the proper times (or affine parameters) of all particles involved in the process, and we denote by
$(\chi_i, \zeta_i)$ and by $(\xi_f,\eta_f)$ the pairs of normalized null five-vectors parameterizing the geodesics of the incoming and outgoing
particles, respectively. Without loss of generality, up to an $SO(1,4)$ transformation, we can choose $X_0$ to be the origin $\bar X=(0,0,0,0,R)$.
We also choose the corresponding normalization constants $k_i$ and $\tilde k_f$ to be the same for all particles. Then we have
\begin{eqnarray}
\zeta_i=\chi_i-\frac {2m_i}{kR} X_0, \ i=1,2;\qquad\ \eta_f=\xi_f-\frac {2\tilde m_f}{kR} X_0,\ f=1,2,\ldots,N,\label{240}
\end{eqnarray}
and
\begin{eqnarray}
K_i=\frac {ck}R X_0\wedge \chi_i, \ i=1,2;\qquad\ \tilde K_f=\frac {kc}{R} X_0\wedge \xi_f, \ f=1,2,\ldots,N,\label{241}
\end{eqnarray}
so that the conservation equation (\ref{238}) reads
\begin{eqnarray}
(\chi_1+\chi_2-\sum_{f=1}^N \xi_f)\wedge X_0=0.\label{242}
\end{eqnarray}
With the choice $X_0=\bar X$ equation (\ref{242}) becomes
\begin{eqnarray}
\chi^\mu_1+\chi^\mu_2=\sum_{f=1}^N \xi^\mu_f,\ \mu=0,1,2,3,\label{243}
\end{eqnarray}
while, using (\ref{240}) (for simplicity we suppress the indices $i$ and $f$)
and since the vectors $\chi$, $\zeta$, $\xi$, $\eta$ are null, we get
\begin{eqnarray}
\chi=(\chi^0,\vec \chi,\frac mk),\quad \zeta=(\chi^0, \vec \chi, -\frac mk),
\quad \xi=(\xi^0,\vec \xi,\frac {\tilde m}k),\quad \eta=(\xi^0, \vec \xi, -\frac {\tilde m}k).\label{246}
\end{eqnarray}
Furthermore, from (\ref{126}) we have
\begin{eqnarray}
\left. m\frac {dX^\mu}{d\tau}\right|_{\tau=0} =kc\chi^\mu =:q^\mu, \quad\ \left. m\frac {dX^4}{d\tau}\right|_{\tau=0}=0, \label{247}
\end{eqnarray}
and a similar formula for $\xi$. Then
\begin{eqnarray}
&& q^2=(q^0)^2-(\vec q)^2=m^2 c^2, \quad\ \tilde q^2=(\tilde q^0)^2-(\vec {\tilde q})^2=\tilde m^2 c^2,\label{248}
\end{eqnarray}
and (\ref{243}) can be rewritten as
\begin{eqnarray}
q_1^\mu +q_2^\mu =\sum_{f=1}^N \tilde q^\mu_f. \label{249}
\end{eqnarray}
In a neighborhood of $X_0=\bar X$ in $dS(1,3)$ we choose local coordinates $x^\mu$, defined by $x^\mu=X^\mu$, $\mu=0,1,2,3$. Since the hyperplane
$X^4=R$ is tangent to $dS(1,3)$ at $X_0$, we have $\left. \partial X^4/ \partial x^\mu \right|_{X_0}=0$ so that,
at $X_0$, the metric of $dS(1,3)$, expressed in terms of the coordinates $x^\mu$, is given by
$ds^2|_{X_0}=(\eta_{AB} dX^A dX^B)|_{dS(1,3),X_0}=\eta_{\mu\nu} dx^\mu dx^\nu$. Then the $x^\mu$ are locally Lorentzian at $X_0$ and
$dX^\mu /d\tau|_{\tau=0}=dx^\mu/d\tau|_{\tau=0}$. Therefore,
\begin{eqnarray}
\left. q^\mu_i=m_i\frac {dx^\mu_i}{d\tau}\right|_{\tau=0} =\pi^\mu_i \qquad (i=1,2)\label{250}
\end{eqnarray}
are the components of the four-momentum of the ingoing particles in the given local frame. Similarly,
\begin{eqnarray}
\tilde q^\mu_f =\tilde \pi^\mu_f \qquad (f=1,2,\ldots,N),\label{251}
\end{eqnarray}
and equation (\ref{249}) can be identified with (\ref{236}).

%%%%%%%%%%%%%%%%%%%%%%%%%%%%%%%%%%%%%%%%%%%%%%%%%%%%%%%%%%
\subsection{Detection}\label{ssec:detection}
In a process such as (\ref{235}), the energies and the momenta of the incoming and outgoing particles are not measured at the interaction point
$X_0$ but at points far away from it. Therefore, we need a formula which relates any given four-momentum $\pi^\mu$ (or $\tilde \pi^\mu$) at $X_0$
to the same parallelly transported four-momentum (along the corresponding geodesic) at the point $X_1$, where it is measured. Since the quantities
$K_{AB}$ are conserved along the geodesic motion, this can be obtained by equating the r.h.s. of (\ref{237}) to the r.h.s. of the same formula
evaluated at $x_1$ ($X_1=X(x_1)$):
\begin{eqnarray}
\left. \left( X_A \frac {\partial X_B}{\partial x^\mu} -X_B \frac {\partial X_A}{\partial x^\mu}\right)\right|_{x=x_0} \pi^\mu (x^0)=
\left. \left( X_A \frac {\partial X_B}{\partial x^\mu} -X_B \frac {\partial X_A}{\partial x^\mu}\right)\right|_{x=x_1} \pi^\mu (x^1).
\label{252}
\end{eqnarray}
Upon multiplying both sides of this equation by
$\left. \left( X_A {\partial X_B}/{\partial x^\mu} -X_B {\partial X_A}/{\partial x^\nu}\right)\right|_{x=x_0}$ and summing over
$A$ and $B$, we find
\begin{eqnarray}
\pi^\mu(x_0)=G^\mu_{\ \ \nu} (x_0,x_1) \pi^\nu (x_1),\label{253}
\end{eqnarray}
where
\begin{eqnarray}
G^\mu_{\ \ \nu} (x_0,x_1)=-\frac 1{2R^2} g^{\mu\rho}(x_0)
\left. \left( X^A \frac {\partial X^B}{\partial x^\rho} -X^B \frac {\partial X^A}{\partial x^\rho}\right)\right|_{x=x_0}
\left. \left( X_A \frac {\partial X_B}{\partial x^\nu} -X_B \frac {\partial X_A}{\partial x^\nu}\right)\right|_{x=x_1}
\label{254}
\end{eqnarray}
with
\begin{eqnarray}
g_{\mu\nu} (x) =\eta_{AB} \frac {\partial X^A}{\partial x^\mu} \frac {\partial X^B}{\partial x^\nu}.\label{255}
\end{eqnarray}
We calculate the matrix $\{ G^\mu_{\ \ \nu} \}$ in flat coordinates and in the case of two-dimensional de Sitter $dS(1,1)$, choosing $x_0=(0,0)$
and $x_1=(ct,x)$. We find
\begin{eqnarray}
\{G^\mu_{\ \ \nu} (x_0,x_1) \}=\left(
\begin{array}{cc}
1 & e^{2\frac {ct}R} \frac xR \\ \frac xR & e^{\frac {ct}R} \cosh \frac {ct}R +e^{2\frac {ct}R} \frac {x^2}{2R^2}
\end{array}
\right).
\end{eqnarray}
The flat coordinates are locally inertial at $x_0$, but not so at $x_1$, because of the scale factor $a(t)=e^{ct/R}$ in the metric (\ref{flatmetric}).
Therefore, if we want to compare the two-momentum $\pi^\mu$ at the points $x_0$ and $x_1$ relative to locally inertial frames at rest
at each such point (with respect to the flat coordinates), we must transform it with the identity matrix at $x_0$ and with the matrix
$$
\left( \begin{array}{cc} 1 & 0 \\ 0 & e^{\frac {ct}R} \end{array} \right)
$$
at $x_1$. Therefore, denoting the inertial energy-momenta at $x_0$ and $x_1$ by $\hat \pi^\mu (x_0)$ and $\hat \pi^\mu (x_1)$, respectively,
we have
\begin{eqnarray}
&& \hat \pi^0 (x_0)\!=\pi^0 (x_0), \quad \hat \pi^1 (x_0)=\!\pi^1 (x_0),\quad
\hat \pi^0 (x_1)\!=\pi^0 (x_1), \quad \hat \pi^1 (x_1)\!=e^{\frac {ct}R}\pi^1 (x_1),\label{257}
\end{eqnarray}
so that
\begin{eqnarray}
&&\!\!\!\!\!\! \hat \pi^0 (x_0)\!=\!\hat \pi^0 (x_1)\! +\! e^{\frac {ct}R} \frac xR \hat \pi^1 (x_1), \quad
\hat \pi^1 (x_0)\!=\!\frac xR \hat \pi^0 (x_1)\! +\!\! \left(\!\cosh \frac {ct}R\! +\! e^{\frac {ct}R} \frac {x^2}{2R^2}\! \right)\! \hat \pi^1 (x_1).
\label{258}
\end{eqnarray}
For example, consider the case of a photon transmitted from $x_0$ to $x_1$ for which
\begin{eqnarray}
\hat \pi^0 (x_0)=\hat \pi^1 (x_0)=\frac {h\nu_0}c, \quad \hat \pi^0 (x_1)=\hat \pi^1 (x_1)=\frac {h\nu}{c}.\label{259}
\end{eqnarray}
The trajectory is a lightlike geodesic with $c^2 dt^2=e^{2ct/R} dx^2$. Hence, up to a sign, we have
\begin{eqnarray}
&& x(t)=R(1-e^{-\frac {ct}R}).\label{260}
\end{eqnarray}
Inserting this into (\ref{258}) and taking (\ref{259}) into account yields the well known formula for the redshift due to the de Sitter expansion
\begin{eqnarray}
\nu=\nu_0 e^{-\frac {ct}R}. \label{261}
\end{eqnarray}
In particular, $\nu(\infty)=0$ which means that the photon becomes black as the comoving emitter approaches the cosmological horizon.

%%%%%%%%%%%%%%%%%%%%%%%%%%%%%%%%%%%%%%%%%%%%%%%%%%%%%%%%%%%%%%%%%%%%%%%%%%%%%%%%%%%%%%%%%%%%%%%%%%%%%%%%%
%%%%%%%%%%%%%%%%%%%%%%%%%%%%%%%%%%%%%%%%%%%%%%%%%%%%%%%%%%%%%%%%%%%%%%%%%%%%%%%%%%%%%%%%%%%%%%%%%%%%%%%%%
\section{Killing vectors in flat Friedmann and de Sitter universes}\label{sec:sasha}
In the preceding sections we investigated several properties of de Sitter spacetime $dS(1,3)$ as the maximally
symmetric four-dimensional Lorentzian manifold embodying the basics symmetries of spacetime homogeneity, isotropy of space and boost
invariance. And we have argued that, due to the observational evidence of the existence of a nonzero cosmological constant $\Lambda$, it
is indeed $dS(1,3)$ which should be considered as the classical background arena for all natural phenomena. On this basis we have assumed the stance
that, at least on foundational grounds, traditional special relativity based on flat Minkowski spacetime should be replaced by
a de Sitter special relativity. In practice, of course, due to the smallness of $\Lambda$, the Minkowskian approximation works
perfectly well at the level of local experiments.\\
However, the de Sitter universe also constitutes a natural bridge between special relativity and general relativity. Indeed, the de Sitter
spacetime can be treated as a member of the family of Friedmann spacetimes \cite{Friedmann}, i.e. spatially homogeneous and isotropic expanding universes.
The peculiarity of the de Sitter cosmology consists in the fact that it belongs simultaneously to all the three families of Friedmann universes:
flat, closed and open, depending on the choice of the coordinate system. Instead, if one considers a quasi--de Sitter cosmology,
for example the cosmology of an isotropic and homogeneous universe in which, together with a cosmological constant, a small amount of dust--like matter
is present, this ``topological degeneracy'' disappears and changing the coordinate system does not affect the spatial topology:
in the presence of matter the overall energy density is critical or larger or smaller than critical. Such a peculiarity of the de Sitter universe is
connected with its maximal symmetry. What happens if the symmetry is reduced, while still keeping, however, the homogeneity and isotropy
of space? In this section we look at this problem from the point of view of differential geometry. Precisely, we consider a family of
flat Friedmann universes with an arbitrary law of dependence of the scale factor on the cosmic time parameter and look for its symmetry
generators by solving the corresponding Killing equations (see appendix C). We prove that any such Friedmann universe either has exactly
six Killing fields, corresponding indeed to invariance under space rotations and translations, or it is maximally symmetric, being
thus de Sitter (or Minkowski). In other words, there are no flat Friedmann metrics which are not maximally symmetric but have more than six independent
Killing fields. Similar results are obtained if one starts from closed or open Friedmann universes instead.
There are just two exceptions. One is the degenerate static Einstein universe \cite{Einstein} which has seven Killing fields
(it is also invariant under time translations) and is spatially closed.
The second one is the hyperbolic analogue of the former: it is spatially open, and has a negative cosmological constant and a negative energy
density.\\
\indent
The results of this section strengthen those of section \ref{sec:spacetime}
since they show that, apart from the exceptional cases of Einstein's static
universe and of its hyperbolic analogue,
boost invariance is implied by spacetime homogeneity and by isotropy of space, without further assumptions. Thus we have a situation
which is similar, though more general, to the one which one encounters in traditional Einsteinian relativity. Once we drop the assumption of
commutativity of spacetime translations, we find that, barring pathological situations, the de Sitter group (together with all its possible contractions)
is the only kinematical group which is compatible with spacetime homogeneity and isotropy of space. {\it This is the content of the principle
of the absolute world in its widest meaning}.\\
\indent We start from the flat Friedmann metric
\begin{equation}
ds^2 = dt^2 -a^2(t)(dx^2 + dy^2 + dz^2) \label{flat}
\end{equation}
in which we have set $c=1$ without loss of generality. We do not make any assumption about the time dependence of the scale factor $a(t)$.
The Killing equations (\ref{B28})
for this metric give the following set of independent equations (the dot denotes the derivative with respect to $t$):
\begin{eqnarray}
&& \frac {\partial \xi_{t}}{\partial t} = 0, \label{Killing1} \quad\
\frac {\partial \xi_{x}}{\partial x} = \dot{a}a\xi{_t}, \quad\
\frac {\partial \xi_{t}}{\partial x} + \frac {\partial \xi_{x}}{\partial t} = 2\frac{\dot{a}}{a}\xi_{x}, \quad\
\frac {\partial \xi_{x}}{\partial y} + \frac {\partial\xi_{y}}{\partial x} = 0,
\end{eqnarray}
plus the ones obtained from these by circular permutations of the spatial indices.
From the first of (\ref{Killing1}) it follows that
\begin{equation}
\xi_{t} = \xi_{t}(x,y,z).\label{xi-t}
\end{equation}
Substituting (\ref{xi-t}) into the second of (\ref{Killing1}) and integrating we get
\begin{equation}
\xi_{x} = \dot{a}a\int\xi_{t}(x,y,z)dx + f(y,z,t),  \label{xi-x}
\end{equation}
which, substituted into the third of (\ref{Killing1}), yields
\begin{equation}
\frac {\partial \xi_{t}}{\partial x} = (\dot{a}^2 - \ddot{a}a)\int\xi_{t}(x,y,z)dx -
\left(\frac {\partial f}{\partial t} - 2\frac{\dot{a}}{a}f\right).\label{xi-t1}
\end{equation}
Taking the time derivative of (\ref{xi-t1}) and taking into account (\ref{xi-t}) gives
\begin{equation}
\left[\frac{d}{dt}(\dot{a}^2-\ddot{a}a)\right]\int\xi_{t}(x,y,z)dx - \frac{\partial}{\partial t}\left(\frac {\partial f}{\partial t} -
2\frac{\dot{a}}{a}f\right) = 0.\label{xi-t2}
\end{equation}
The two terms at the l.h.s. of (\ref{xi-t2}) should vanish separately, because the first term
depends on $x$ while the second does not.
Thus, we have
\begin{eqnarray}
\left[\frac{d}{d t}(\dot{a}^2-\ddot{a}a)\right]\int\xi_{t}(x,y,z)dx = 0,
\qquad\ \mbox{ and } \qquad\
\frac{\partial}{\partial t}\left(\frac {\partial f}{\partial t} -2\frac{\dot{a}}{a}f\right) = 0.
\end{eqnarray}
If
\begin{equation}
\frac{d}{d t}(\dot{a}^2-\ddot{a}a) \neq 0, \label{condition3}
\end{equation}
we have
\begin{eqnarray}
&&\xi_{t} = 0, \quad\ \frac {\partial \xi_x}{\partial x} = 0
\qquad\ \mbox{ and } \qquad\
\frac {\partial \xi_{x}}{\partial t} = 2\frac{\dot{a}}{a}\xi_{x}. \label{conseq2}
\end{eqnarray}
Then, integrating the third of the equations (\ref{conseq2}) (and the corresponding equations for $\xi_y$ and $\xi_z$) we obtain
\begin{eqnarray}
\xi_{x} = {C(y,z)}{a^2}, \quad\ \xi_{y} = {D(z,x)}{a^2}, \quad\ \xi_{z} = {E(x,y)}{a^2}. \label{xi-z1}
\end{eqnarray}
Substituting (\ref{xi-z1}) into the fourth equation of
(\ref{Killing1}) and into the analogous equations obtained by permutation of the spatial indices, we get
\begin{equation}
\frac {\partial C}{\partial y} +\frac {\partial D}{\partial x}=\frac {\partial D}{\partial z}+\frac {\partial E}{\partial y}=
\frac {\partial E}{\partial x} +\frac {\partial C}{\partial z}  = 0.\label{condition4}
\end{equation}
Differentiating (\ref{condition4}) yields
\begin{equation}
\frac {\partial^2 D}{\partial x^2}=\frac {\partial^2 D}{\partial z^2}=\frac {\partial^2 C}{\partial y^2}
=\frac {\partial^2 C}{\partial z^2}=\frac {\partial^2 E}{\partial x^2}=\frac {\partial^2 E}{\partial y^2} = 0 \label{D}
\end{equation}
and
\begin{eqnarray}
\frac {\partial^2 C}{\partial y \partial z}+\frac {\partial^2 D}{\partial x \partial z}=0, \quad
\frac {\partial^2 D}{\partial z \partial x}+\frac {\partial^2 E}{\partial y \partial x}=0, \quad
\frac {\partial^2 E}{\partial x \partial y}+\frac {\partial^2 C}{\partial z \partial y}=0. \label{nome}
\end{eqnarray}
Combining together relations (\ref{nome}) gives
\begin{eqnarray}
\frac {\partial C}{\partial y \partial z}=\frac {\partial D}{\partial z \partial x}=\frac {\partial E}{\partial x \partial y}=0.
\end{eqnarray}
Then, the contravariant components $\xi^\mu (x)$ of the Killing vector $\xi$ are of the form
\begin{eqnarray}
\xi^0=0, \quad \xi^1=\alpha+\beta y +\gamma z, \quad \xi^2=\tilde \alpha +\tilde \beta x +\tilde \gamma z, \quad
\xi^3=\hat \alpha+\hat \beta x +\hat \gamma y.
\end{eqnarray}
Then, using again the fourth of (\ref{Killing1}) and the like, we find that the general Killing field $\xi^\mu (x) \partial/\partial x^\mu$
is a linear combination of the independent fields
\begin{eqnarray}
&&\frac{\partial}{\partial x},\ \frac{\partial}{\partial y},\ \frac{\partial}{\partial z}, \
x\frac{\partial}{\partial y} - y\frac{\partial}{\partial x},\ y\frac{\partial}{\partial z} - z\frac{\partial}{\partial y},\
z\frac{\partial}{\partial x} - x\frac{\partial}{\partial z}, \label{rot}
\end{eqnarray}
which are the standard translation and rotation operators.
Then we conclude that if the scale factor $a(t)$ satisfies condition (\ref{condition3}), the metric (\ref{flat}) has no independent
Killing vectors other than (\ref{rot}).\\
\indent
The option
\begin{equation}
\dot{a}^2 = \ddot{a}a\label{option}
\end{equation}
is equivalent to
\begin{equation}
\frac{\dot{a}}{a} = \frac 1R = const, \label{option1}
\end{equation}
and the constant can be chosen to be positive. Therefore, up to normalization of the time coordinate, we have
\begin{equation}
a(t) = e^{\frac tR}, \label{Hubble}
\end{equation}
and the metric becomes
\begin{eqnarray}
ds^2=dt^2-e^{2\frac tR} (dx^2+dy^2+dz^2),
\end{eqnarray}
which is de Sitter in flat coordinates.\\
\indent Finally, we examine the option
\begin{eqnarray}
\label{321} \dot a^2-\ddot a a =A_0 =const \neq 0.
\end{eqnarray}
Differentiating the third equation of (\ref{Killing1}) w.r.t. $x$ and using the first and second of (\ref{Killing1}) and (\ref{321}), we get
\begin{eqnarray}
\label{322} \frac {\partial^2 \xi_t}{\partial x^2} -A_0 \xi_t =0
\end{eqnarray}
and, similarly,
\begin{eqnarray}
&& \frac {\partial^2 \xi_t}{\partial y^2} -A_0 \xi_t =0, \quad\ \frac {\partial^2 \xi_t}{\partial z^2} -A_0 \xi_t =0. \label{324}
\end{eqnarray}
The general solution of eqs. (\ref{322})--(\ref{324}) is
\begin{eqnarray}
\label{325} \xi_t=\left(A_1 e^{\sqrt {A_0} x} +A_2 e^{-\sqrt {A_0} x}\right)\left(B_1 e^{\sqrt {A_0} y} +B_2 e^{-\sqrt {A_0} y}\right)
\left(C_1 e^{\sqrt {A_0} z} +C_2 e^{-\sqrt {A_0} z}\right).
\end{eqnarray}
Substituting (\ref{325}) in (\ref{xi-x}) (and in the corresponding expression for $\xi_y$)
\begin{eqnarray*}
\!\!\!\!\!\!\xi_x\!=\!\frac {\dot a a}{\sqrt {A_0}}\! \left(\!A_1 e^{\sqrt {A_0} x}\! -\!A_2 e^{-\sqrt {A_0} x}\!\right)
\!\!\left(\!B_1 e^{\sqrt {A_0} y}
\!+\!B_2 e^{-\sqrt {A_0} y}\!\right)\!\! \left(\!C_1 e^{\sqrt {A_0} z} \!+\!C_2 e^{-\sqrt {A_0} z}\!\right)\!+\!f(y,z,t), \cr
\!\!\!\!\!\!\xi_y\!=\!\frac {\dot a a}{\sqrt {A_0}}\! \left(\!A_1 e^{\sqrt {A_0} x}\! +\!A_2 e^{-\sqrt {A_0} x}\!\right)
\!\!\left(\!B_1 e^{\sqrt {A_0} y}
\!-\!B_2 e^{-\sqrt {A_0} y}\!\right)\!\! \left(\!C_1 e^{\sqrt {A_0} z} \!+\!C_2 e^{-\sqrt {A_0} z}\!\right)\!+\!g(z,x,t),\\
\end{eqnarray*}
which inserted into the last of (\ref{Killing1}) yields
\begin{eqnarray}
\label{328}&& 2\dot a a \left(A_1 e^{\sqrt {A_0} x} -A_2 e^{-\sqrt {A_0} x}\right)\left(B_1 e^{\sqrt {A_0} y}
-B_2 e^{-\sqrt {A_0} y}\right) \left(C_1 e^{\sqrt {A_0} z} +C_2 e^{-\sqrt {A_0} z}\right)\cr
&& \qquad +\frac {\partial f(y,z,t)}{\partial y}+\frac {\partial g(z,x,t)}{\partial x}=0.
\end{eqnarray}
The first term at the l.h.s. of (\ref{328}) is a product of functions of $x$, $y$ and $z$ respectively, hence it cannot be compensated
by the second and the third terms. Therefore, (\ref{328}) can only be satisfied if $\xi_t=0$, which leads us back to the first case. In conclusion,
option (\ref{321}) implies once more that there are only six independent Killing fields which can be chosen to be specified by (\ref{rot}).

%%%%%%%%%%%%%%%%%%%%%%%%%%%%%%%%%%%%%%%%%%%%%%%%%%%%%%%%%%%%%%%%%
\section{Conclusions}\label{conclusions}
In the preceding sections
we have developed, at a purely classical level, several introductory and elementary aspects of a theory of special relativity
based on $SO(1,4)$. \\
\indent The idea that special relativity could be founded on de Sitter space rather than on Minkowski space appeared {\it in nuce} long ago, starting perhaps first with a remarkable paper by Dirac \cite{dirac} in which the author puts forward the view that relativistic wave equations for elementary particles should be de Sitter invariant rather than Poincar\'e invariant. For some further literature on the subject see e.g. \cite{dirac,sch1,sch2,gursey,fro,roman,nact}. However, to our knowledge, at the level of a somewhat systematic treatment, investigations of de Sitter relativity have been carried out by two other groups \cite{Al,Guo}. The authors in \cite{Al} employ stereographic coordinates to parameterize $dS(1,3)$. Using these
coordinates they are able to represent the translation generators $P_\mu=N_{\mu 4}/R$ in (\ref{formule}) ($\mu=0,1,2,3$) as a linear
combination of ordinary translations $\Pi_\mu$ plus proper conformal transformations $K_\mu$. The relative importance of these
two terms in the combination depends on the value of $\Lambda$. In particular, for very large values of $\Lambda$ the dominant contribution
comes from the $K_\mu$'s and spacetime assumes a conical structure. The authors claim that this situation may be relevant at the time of
inflation, when the effective cosmological constant was very large ($\Lambda \simeq 10^{56} cm^{-2}$). Instead, in \cite{Guo}
an approach is put forward based on the covering of de Sitter space by Beltrami coordinate patches. It turns out that, relative to these coordinates,
geodesics are represented by straight worldlines. This property leads the authors to claim that the choice of the Beltrami
coordinates is the most suitable one to express the law of inertia in de Sitter relativity. \\
\indent If one would wish to go beyond the elementary treatment of de Sitter relativity given here, the next step would be to develop
a field theory on the de Sitter background, both classical and quantum. This subject is beyond the scope of this paper. Therefore,
we shall confine ourselves here to a few remarks concerning the quantum theory on a de Sitter background. This subject is a particular
aspect of the more general problem of formulating a quantum field theory on an arbitrary curved spacetime \cite{Birrel}. Compared to the
general situation, things are simpler in the de Sitter case because of maximal symmetry. However, also in this case, things are complicated
by the fact that the notion of a particle as an excitation of a certain quantum field propagating in the spacetime manifold
becomes highly ambiguous because of the lack of an asymptotic condition \cite{Birrel,M}. This problem reflects itself in the ambiguity
of providing a well defined characterization of a ground state or ``vacuum'' of the field. In turn, this is linked to the lack of
a global energy operator and, therefore, to the impossibility of formulating a consistent spectral condition \cite{BEM}. \\
\indent Recently, an interesting attempt to overcome this problem has been put forward \cite{BEM,BGM}. It is based on a formulation of the
Wightman axioms for a general scalar field on $dS(1,3)$ in which the spectral condition is replaced by suitable global analyticity
properties of the Wightman functions in the complexified de Sitter manifold. They closely mirror the analogous
analyticity properties which provide an equivalent characterization of the ordinary spectral condition in the Minkowskian case.
The ground states of this (possibly interacting) field theory are exactly those which
satisfy the KMS condition \cite{KMS} with respect to the time translation groups of the geodesic observers of $dS(1,3)$ corresponding to the
temperature $T=\hbar c/(2\pi k R)$. This result provides a rigorous foundation to the discovery by Gibbons and Hawking \cite{Gibbons}
that a geodesic observer in $dS(1,3)$, equipped with a particle detector, will observe a thermal radiation at the given temperature, coming
apparently from the observer's cosmological horizon. It is well known that this effect is the analogue in $dS(1,3)$ of the Hawking effect of
thermal black hole evaporation \cite{Hawking} and of the thermal excitation of a uniformly accelerated detector in Minkowski space \cite{bib}.
In \cite{GKCM} we have discussed a classical analogue of the thermal nature of the de Sitter ``vacua''.\\
\indent However, we must mention that,
as far as interacting fields are concerned, practical calculations of particle scattering and decay amplitudes in a de Sitter background are still at
a very early stage. It is true that, due to the smallness of the cosmological constant, a development of a quantum theory of
interacting fields in dS(1,3) (such as
a de Sitter QED or QCD) is more of foundational than of practical importance for particle physics at the present epoch. It may, however,
be important for the analysis of particle processes at the epoch of inflation, when the effective cosmological constant was very large
\cite{Linde}. For example, a possible theoretical understanding of the structure of the universe as it is today may be based on the nature
of the fluctuations of the quantum fields on the inflationary de Sitter spacetime, fluctuations which are held responsible for
the primordial density inhomogeneities which
have given origin to the structures which exist in the universe today.\\
\indent Perturbative calculations of transition amplitudes in $dS(1,3)$ are much more complicated than in the Minkowski case due to the lack
of commutativity of spacetime translations \cite{BGM,BEM}. To our knowledge, so far the only perturbative problem in the context of an
explicit model of interacting quantum fields, which has been analyzed in detail in a de Sitter background, concerns the computation to lowest order
in the coupling constant of the probability per unit time of a scalar particle of mass $m$ to decay into two identical scalar particles
of mass $m_1$ \cite{BEM}. Contrary to what happens in the flat case, it turns out that, provided $m$ is larger than a critical value $m_c$,
the decay probability is nonzero, even if $m<2m_1$. From a group theoretical point of view this result is a consequence of the fact that the
direct integral decomposition into irreducible components of the direct product by itself of the irreducible representation of $SO(1,4)$
corresponding to mass $m_1$ contains contributions from all masses $m>m_c$. A more intuitive physical interpretation of the result is the
following: if $m<2m_1$, in Minkowski space the decay process $m\rightarrow 2m_1$ is forbidden by energy conservation.
In a de Sitter background it is also forbidden classically (see sec. \ref{sec:kinematics}) since the process is strictly localized,
hence it does not feel the repulsive effect of the cosmological constant.
On the other hand, quantum mechanically, the spread of the wave packet is sensitive to this effect, which can therefore provide the
missing energy which is needed for the process to take place.
In particular, if one computes the decay probability to lowest order in the cosmological constant, one finds that
the correction to the flat term is proportional to $e^{-|\Delta m|/m_{dS}}$, where $\Delta m=2m_1-m$ and $m_{dS}=\hbar/{Rc}$ is the
de Sitter mass. Since $R\simeq 10^{28} cm$ we have $m_{dS}\simeq 10^{-65} g$ ($\simeq 10^{-32} eV$) so that no reasonable process of this
kind can take place at the present epoch. On the other hand, processes of this type may have been frequent during the inflation era, when $R$
was much smaller.
The preceding considerations indicate that a consistent study of quantum field theory in a de Sitter background is desirable due to its
potentially important impact in particle physics and cosmology.\\
\indent We conclude with the following lyrically sad observation. If dark energy indeed corresponds to a cosmological constant, which is the feeling that we have,
the de Sitter geometry is the one which the geometry of the universe does asymptotically tend to. Then, in spite
of its great importance for cosmology, it is perhaps an ironical fact that the cosmological constant will also eventually bring
cosmology to an end \cite{Krauss}.

\begin{acknowledgement}
We are grateful to Luca Rizzi for the skillful processing of the figures.
\end{acknowledgement}

\begin{appendix}
%%%%%%%%%%%%%%%%%%%%%%%%%%%%%%%%%%%%%%%%%%%%%%%%%%%%%%%%%%%%%
%%%%%%%%%%%%%%%%%%%%%%%%%%%%%%%%%%%%%%%%%%%%%%%%%%%%%%%%%%%%%
\section{Construction of the de Sitter manifold \boldmath{$dS(1,3)$}}\label{app:manifold}
A real simple Lie group $G$ is naturally provided with a non degenerate metric, the Killing metric. This is obtained starting
from the Killing form on the corresponding Lie algebra ${\cal L}=Lie(G)$:
\begin{eqnarray}
\langle \ , \rangle : {\cal L}\times {\cal L} \longrightarrow  \bR, \quad \ (A,B) \longmapsto \langle A,B\rangle ={\rm Tr}[ ({\rm ad}A)({\rm
ad}B)],\label{scalar}
\end{eqnarray}
where $A\rightarrow {\rm ad} A =[A,\cdot ]$ is the adjoint representation. The next step is to use the fact that the group is
endowed with the left translation
\begin{eqnarray}
L_g:G \longrightarrow G \quad\
h \longmapsto L_g(h)=gh
\end{eqnarray}
which, by the definition of a Lie group, is a differentiable map. This can be used to ``translate'' the Killing form everywhere on $G$ via
pullback. Indeed, $L\simeq T_eG$, $e$ being the unit element, so that one can define the product in $g$
\begin{eqnarray}
ds^2=\langle \ , \rangle_g: T_g G\times T_g G \longrightarrow \bR
\qquad\ \mbox{ as } \qquad\
\langle \ , \rangle_g= L_{g^{-1}}^* \langle \ , \rangle_e=\langle \ , \rangle.
\end{eqnarray}
Let us see how this can be computed practically. Let $\gamma(x)$ be some differentiable local coordinatization of the group:
\begin{eqnarray}
\gamma: U\longrightarrow G.
\end{eqnarray}
An infinitesimal displacement $dx$ in $x\in U$ determines an infinitesimal displacement $dg=d\gamma (x)$ living in $T_g G$,
$g=\gamma(x)$. The left translation can then be used to translate such a displacement in $T_e G$:
\begin{eqnarray}
L_{g^{-1}*} dg =:g^{-1} dg=\gamma(x)^{-1} d\gamma(x).
\end{eqnarray}
Note that this formula holds even if $\gamma$ parameterizes only a submanifold of $G$. In this case this gives the metric induced on
the submanifold
\begin{eqnarray}
ds^2=\langle dg , dg \rangle_g= \langle \gamma(x)^{-1} d\gamma(x) ,
\gamma(x)^{-1} d\gamma(x) \rangle.
\end{eqnarray}
In general, for a simple Lie group, in place of the adjoint representation any faithful representation can be used to compute
the Killing product, all differing by some constant. We use this property to construct the metric on $dS(1,3)$, the hyperboloid
(\ref{hyperboloid}). With this aim, we need to identify the hyperboloid with a submanifold of the group itself and then need to induce the
Killing metric on it. This is done by noting that the natural action of $SO(1,4)$ on itself or on its Lie algebra is the adjoint action
so that, with the identifications $G=SO(1,4)$, $g=H$, $h=L$, we have the actions
\begin{eqnarray}
&& X\longmapsto LX,\\
&& H\longmapsto LHL^{-1},\\
&& H^{-1} dH \longmapsto (LHL^{-1})^{-1} d(LHL^{-1})= (LH^{-1}
L^{-1}) LdH L^{-1} = L H^{-1} dH L^{-1},
\end{eqnarray}
where $X\in dS(1,3)$ and $H, L\in SO(1,4)$. Such actions correspond to isometries of the Killing metric because
\begin{eqnarray}
{\rm Tr} [(L H^{-1} dH L^{-1})(L H^{-1} dH L^{-1})]={\rm Tr}[(H^{-1}
dH )(H^{-1} dH)].
\end{eqnarray}
The subset $S$ of $SO(1,4)$ corresponding to $dS(1,3)$ must then be parameterized by the points $X\in~dS(1,3)$. Therefore, we need to
construct matrices $H(X)$ which transform as $L H(X) L^{-1}$ under a transformation $X\rightarrow LX$. Thus $H(X)$ can be thought of as a
tensor with a covariant index and a contravariant index. The most general tensor of this kind that we can construct with $X$ is
\begin{eqnarray}
H^A_{\ B}= \alpha \delta^A_B +\beta X^A X_B,
\end{eqnarray}
with $\alpha$ and $\beta$ as scalar coefficients. To live in $SO(1,4)$, $H$ must satisfy
\begin{eqnarray}
\eta_{AB}=H^C_{\ A}H^D_{\ B} \eta_{CD},
\end{eqnarray}
which yields $\alpha=\pm 1$ and $\beta=\pm2/R^2$. The condition ${\rm det} H=1$ forces to take the minus sign. Thus,
\begin{eqnarray}
H^A_{\ B}(X)=-\delta^A_{\ B}-\frac 2{R^2} X^AX_B.
\end{eqnarray}
Note that $H^2=\bb {I}$ so that $H=H^{-1}$ and $H dH+dH H=0$. Then
\begin{eqnarray}
{\rm Tr} (H^{-1}dH H^{-1}dH)={\rm Tr} (H dH H^{-1}dH)=-{\rm Tr} (dH H H^{-1}dH)=-{\rm Tr} (dH dH).
\end{eqnarray}
Thus we find
\begin{eqnarray}
ds^2=-\frac 4{R^4} {\rm Tr} (d(X^A X_C)d(X^C X_B))=\frac 8{R^2} \eta_{AB} dX^A dX^B|_{X^2=-R^2}.
\end{eqnarray}
Dropping the unessential constant factor $8/R^2$, we see that the invariant metric on $dS(1,3)$ is exactly the one inherited from its
embedding in $M(1,4)$.

%%%%%%%%%%%%%%%%%%%%%%%%%%%%%%%%%%%%%%%%%%%%
\section{Curvature tensor} \label{app:curvatura}
In this appendix we collect some technical facts about geometry on curved manifolds from an elementary point of view.
%%%%%%%%%%%%%%%%%%%%%%%%%%%%%%%%
\subsection{Covariant derivatives, connection}\label{sapp:covderiv}
A non trivial problem on curved manifolds $M$ is the calculation of derivatives of vector fields. Usually, a derivative compares the
values of a function at points which are close to each other. Now, a vector field $v$
is a function that at any point $p\in M$ associates a vector $v(p)$ in the tangent space $T_p M$ at $M$ in $p$. Then, if $p\neq q$ are two distinct
points of $M$, $v(p)$ and $v(q)$ are vectors belonging to different tangent
spaces and cannot be directly compared. Therefore, to compare $v(q)$ and $v(p)$ we need means to transport, say,
$v(p)$ to the tangent space at $q$. This operation is called {\it parallel transport}. Then, suppose we move from $p$ to $q$ along a curve
\begin{eqnarray}
\gamma: [0,h] \longrightarrow M\ , \qquad \ \gamma(0)=p\ , \quad \gamma(h)=q \ .\label{B1}
\end{eqnarray}
For an infinitesimal displacement ($h\rightarrow 0$) we expect that under the transport from $p$ to $q$ the vector $v(p)$ will change by
an infinitesimal amount, proportional to the displacement:
$v(p)\mapsto v(p)+\delta v$. Obviously $\delta v$ must also depend linearly on $v(p)$. Then, if $x^\mu$, $\mu=0,\ldots,3$, are local
coordinates in a neighborhood containing $p$ and $q$ and if we put
$\delta x^\mu =x^\mu(\gamma(h))-x^\mu(\gamma(0))$, the infinitesimal change, expressed in terms of the corresponding
contravariant components of the vector field, will be
\begin{eqnarray}
v^\mu (p) \longmapsto \tau_q(v(p)):=v^\mu (p)+\delta v^\mu= v^\mu(p)- \Gamma^\mu_{\rho
\sigma}(p) \delta x^\rho v^\sigma (p) \ ,\label{B2}
\end{eqnarray}
where we use the Einstein convention: repeated indexes are summed over. The coefficients $\Gamma^\mu_{\rho \sigma}(p)$, which depend on $p$,
define the {\it connection}. The sign is conventional. Now the transported vector $\tau_q (v(p))$ belongs to the same tangent space
as $v(q)$ so that we can subtract it from $v(q)$ and define, in the limit $h\rightarrow 0$, the {\it covariant derivative } $\nabla_\mu v (p)$ as
\begin{eqnarray}
\delta x^\mu \nabla_\mu v (p) := v(q) -\tau_q (v(p)).\label{B3}
\end{eqnarray}
Thus, in terms of components,
\begin{eqnarray}
\nabla_\mu v^\nu = \frac {\partial v^\nu}{\partial x^\mu} +\Gamma^\nu_{\mu\sigma} v^\sigma \ .\label{B4}
\end{eqnarray}
Formula (\ref{B4}) defines the derivative of a contravariant vector field. To establish the covariant derivative of a covariant vector
field $w_\mu$ (one-form) we proceed as follows. The expression $w_\mu v^\mu$ is a scalar, hence it does not change under transport and its
covariant derivative is its ordinary derivative. Then, requiring the covariant derivative to obey the Leibniz rule, we have
\begin{eqnarray}
&& \frac {\partial w_\nu}{\partial x^\mu} v^\nu+w_\nu \frac {\partial v^\nu}{\partial x^\mu}=\frac {\partial}{\partial x^\mu} (w_\nu v^\nu)
=\nabla_\mu (w_\nu v^\nu)=(\nabla_\mu w_\nu) v^\nu +w_\nu (\nabla_\mu v^\nu)\cr
&&\qquad\ =(\nabla_\mu w_\nu) v^\nu +w_\lambda \left( \frac {\partial v^\lambda}{\partial x^\mu}+\Gamma^\lambda_{\mu\nu} v^\nu \right)
\label{B5}
\end{eqnarray}
so that
\begin{eqnarray}
\nabla_\mu w_\nu =\frac {\partial w_\nu}{\partial x^\mu} -\Gamma^\lambda_{\mu\nu} w_\lambda.\label{B6}
\end{eqnarray}

%%%%%%%%%%%%%%%%%%%%%%%%%%
\subsection{Levi-Civita (Christoffel) connection}\label{sapp:partrans}
In general there is no further rule to determine the coefficients of the connection. However, in our case there are two facts which
give rise to a natural choice of the connection. \\ \indent
The first one is suggested by our discussion in section \ref{ssec:causal} of the geometrical meaning of relations (\ref{101}) and (\ref{103})
expressing the noncommutativity of spacetime translations. We stated there that an infinitesimal displacement in the direction
$i$ composed with an analogous displacement in the direction $j$, followed next by the inverse displacements produces the same result as
a small rotation (or boost) in the $i$--$j$ plane at the starting point.
If we want to realize this fact geometrically, it is natural to ask this sequence of four infinitesimal displacements to lead us back to the starting point. In other
words, the composition of these displacements will define a
closed rectangle. Then, if we denote by $y^\mu$ and $z^\nu$ the two infinitesimal vectors representing such displacements on the given plane,
this rectangle must be generated by first transporting $y^\mu$ from
$x^\mu$ to $x^\mu+z^\mu$ and next transporting $z^\mu$ from $x^\mu$ to $x^\mu+y^\mu$, $x^\mu$ being the coordinates of the starting point $p$. For
simplicity we can assume $x^\mu(p)=0$. The resulting rectangle is closed if $z^\mu+\tau_z (y^\mu)=y^\mu
+\tau_y (z^\mu)$. Using the expression above in terms of the coefficients of the connection, we find
\begin{eqnarray}
z^\mu +y^\mu -\Gamma^\mu_{\sigma\rho} z^\sigma y^\rho =y^\mu +z^\mu -\Gamma^\mu_{\sigma\rho} y^\sigma z^\rho \ , \label{B7}
\end{eqnarray}
which must be true for arbitrary $y^\mu$ and $z^\nu$. This implies the connection to be symmetric:
\begin{eqnarray}
\Gamma^\mu_{\sigma\rho}=\Gamma^\mu_{\rho\sigma} \ .\label{B8}
\end{eqnarray}
A connection satisfying this property is said to be {\it torsion free }.\\ \indent
However, the torsionless condition is not necessarily the natural one, and theories with torsion appear frequently in physics. See e.g.
\cite{hehl} and references therein.
The vanishing of the torsion is a strong condition which requires further physical motivations.
But here there is no question of invoking such principle since we confine ourselves to formulating a special theory
of relativity. In this perspective, the torsion free condition is just a choice and cannot be deduced. In other words, the naive argument of
the closure of the rectangle should be intended here as the one corresponding to the simplest choice, not as a proof.
A correct presentation for the torsion tensor field is to define it as the vector bundle valued two-form $T$ which,
applied to two vector fields $u$ and $v$, measures the difference between their covariant commutator and their Lie bracket:
\begin{eqnarray}
T(u,v)=\nabla_u v-\nabla_v u -[u,v].
\end{eqnarray}
Here $\nabla_u v$ denotes the covariant derivative of the vector field $v$ along the direction $u$.
See figure 5 for a pictorial representation.

\begin{figure}[h]
\begin{center}
\includegraphics[scale=0.5]{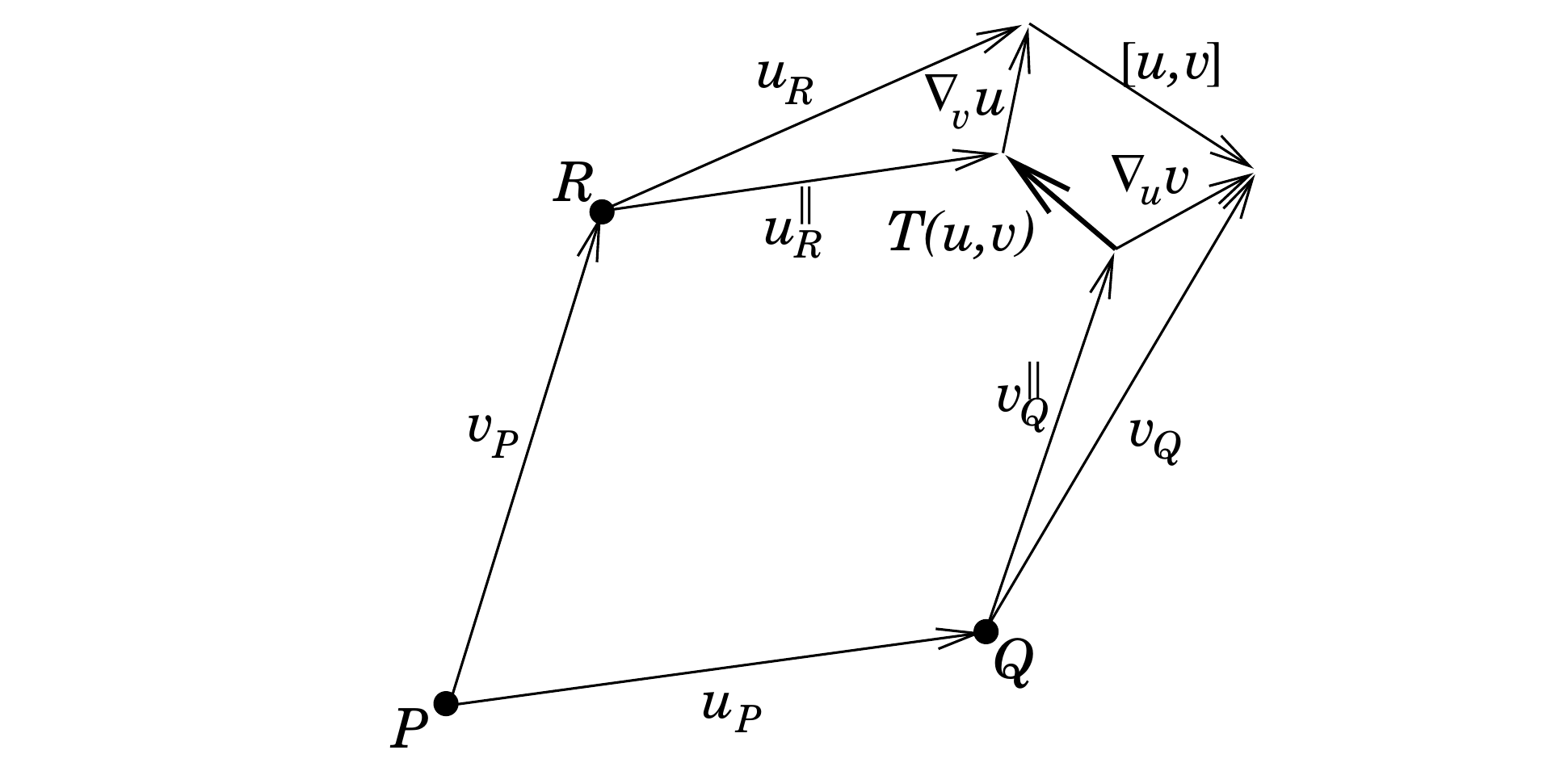}
\caption{\em The torsion tensor.}
\end{center}
\end{figure}

The second property that we require for the connection is related to the presence of a metric on the manifold
\begin{eqnarray}
ds^2 =g_{\mu\nu} dx^\mu dx^\nu \ .\label{B9}
\end{eqnarray}
It is natural to require that the transport should respect the metric. In other words, if $v(p)$ and $w(p)$ are any
two vectors at $p$, their scalar product $g_{\mu\nu} v^\mu w^\nu$ should not change under transport from $p$ to $q$
along any curve. A transport with this property is called a {\it metric-compatible transport}. Then, if $x^\mu$ and
$x^\mu+z^\mu$ are the coordinates of $p$ and $q$ and if $z^\mu$ is small, the parallel transport is determined by the condition
\begin{eqnarray}
g_{\mu\nu}(x) v^\mu w^\nu =(v^\mu -\Gamma^\mu_{\rho\sigma}z^\rho v^\sigma)(w^\nu -\Gamma^\nu_{\alpha\beta}z^\alpha w^\beta)
g_{\mu\nu}(x+z) \ .\label{B10}
\end{eqnarray}
Since $v$, $w$ and $z$ are arbitrary vectors, this relation gives, to first order in $z$,
\begin{eqnarray}
0=\frac {\partial g_{\mu\nu}}{\partial x^\rho}
-\Gamma^\alpha_{\rho\mu} g_{\nu\alpha} -\Gamma^\alpha_{\rho\nu} g_{\mu\alpha} \ .\label{B11}
\end{eqnarray}
Using the symmetry of the connection coefficients and denoting as usual by $g^{\mu\nu}$ the inverse metric of $g_{\mu\nu}$
we can solve this w.r.t. the $\Gamma^\mu_{\nu\rho}$ and find
\begin{eqnarray}
\Gamma^\mu_{\nu\rho}=\frac 12 g^{\mu\lambda}\left(\frac {\partial g_{\lambda\rho}}{\partial x^\nu}
+\frac {\partial g_{\lambda\nu}}{\partial x^\rho}  -\frac {\partial g_{\nu\rho}}{\partial x^\lambda}\right) \ . \label{B12}
\end{eqnarray}
This is called the {\it Levi-Civita (Christoffel) connection}.
It is the unique torsion free connection for which the transport preserves the scalar product of vectors.\\
\indent
Here we also recall that, as is well known, the metric tensor $g_{\mu\nu}$ and its contravariant form $g^{\mu\nu}$
can be used to lower and raise indices of vectors and tensors. For example, $v_\rho=g_{\rho\mu} v^\mu$ and
$v^\sigma =g^{\sigma\nu} v_\nu$.

%%%%%%%%%%%%%%%%%%%%%%%%%%%%%%%%
\subsection{Geodesic motions}\label{geodesicmotion}
If we identify $M$ with the spacetime manifold $W$, the parallel transport of Levi-Civita gives rise to a natural motion of free
particles on $W$. Specifically, suppose a massive particle is moving along a timelike path $\gamma: t\rightarrow \gamma(t)$,
$\gamma(t)\in W$, where $t$ is the time coordinate relative to some local coordinate patch $x^\mu=(ct,\vec x)$.
We denote by $u^\mu=dx^\mu/d\tau$ the four-velocity of the particle, $d\tau =(1/c) ds=(1/c) \sqrt {g_{\mu\nu} (dx^\mu/dt)(dx^\nu/dt)} dt$
being the proper time. The vector $u^\mu$ is tangent to the trajectory $\gamma^\mu (\tau)=x^\mu (\gamma (\tau))$. If the particle is not
acted upon by any force, so that it moves freely, we expect that the parallelly transported vector $u^\mu$ coincides all the time with
$u^\mu$ itself. In other words, the covariant derivative of $u^\mu$ along $\gamma$ should be zero:
\begin{eqnarray}
\frac {d^2\gamma^\mu}{d\tau^2} +\Gamma^\mu_{\sigma\rho} \frac {d\gamma^\sigma}{d\tau} \frac {d\gamma^\rho}{d\tau}=0 \ . \label{B13}
\end{eqnarray}
If we think of $W$ as being uniformly filled with an ideal gas of freely moving particles, then $u^\mu$ is the four-velocity field of the gas
and (\ref{B13}) can be written as
\begin{eqnarray}
u^\nu \nabla_\nu u^\mu =0.\label{B14}
\end{eqnarray}
(\ref{B13}) is called the geodesic equation and its solutions are the geodesic curves.
The geodesics are the stationary curves of the proper time functional
\begin{eqnarray}
\tau[\gamma]=\int \sqrt{g_{\sigma\rho}(\gamma(t))\frac {d\gamma^\sigma}{dt} \frac {d\gamma^\rho}{dt}}dt \ .\label{B15}
\end{eqnarray}

%%%%%%%%%%%%%%%%%%%%%%%%%%%%%%%%%%%%%%%
\subsection{Curvature tensor}\label{sapp:curvature}
It is intuitively clear that the curvature properties of a manifold should be determined by how a vector changes under parallel transport.
And, indeed, these properties are locally characterized completely by a fourth order tensor field, called the {\it curvature tensor},
which determines how a vector changes when it is parallelly transported along an infinitesimally small closed loop. Denoting by $\gamma$
such a small loop we have
\begin{eqnarray}
\oint_\gamma \delta v^\mu =-\oint_\gamma \Gamma^\mu_{\rho\nu} v^\nu dx^\rho \ .\label{B16}
\end{eqnarray}
and using Stokes' theorem to transform the line integral into a surface integral
\begin{eqnarray}
-\oint_\gamma \Gamma^\mu_{\rho\nu} v^\nu dx^\rho=-\frac 12 \int_A
\left[\frac {\partial(\Gamma^\mu_{\rho\nu}v^\nu)}{\partial x^\lambda} -\frac {\partial
(\Gamma^\mu_{\lambda\nu}v^\nu)}{\partial x^\rho}\right]d\sigma^{\lambda\rho} \ , \label{B17}
\end{eqnarray}
where $A$ is the surface bounded by $\gamma$ and  $d\sigma^{\lambda\rho}$ is the surface element. Since the loop is infinitesimal,
we can approximate the integrand with its value at the starting point $x=x_0$, and then we find, up to higher orders,
\begin{eqnarray}
\oint_\gamma \delta v^\mu\simeq -\frac 12 \Sigma^{\lambda\rho}
\left[\frac {\partial(\Gamma^\mu_{\rho\nu}v^\nu)}{\partial x^\lambda} -\frac{\partial
(\Gamma^\mu_{\lambda\nu}v^\nu)}{\partial x^\rho}\right]_{x=x_0} \ ,\label{B18}
\end{eqnarray}
where $\Sigma^{\mu\nu}$ is the area of the projection of the surface
on the $\mu$--$\nu$ plane. Since $v$ is parallelly transported, we have $\frac {\partial v^\nu}{\partial x^\lambda} =-\Gamma^\nu_{\lambda\sigma}
v^\sigma$ so that we finally obtain
\begin{eqnarray}
\oint_\gamma \delta v^\mu= -\frac 12 \Sigma^{\lambda\rho} {\RR^\mu}_{\nu\lambda\rho} v^\nu \ ,\label{B19}
\end{eqnarray}
where
\begin{eqnarray}
{\RR^\mu}_{\nu\lambda\rho}=\frac {\partial
\Gamma^\mu_{\nu\rho}}{\partial x^\lambda}-\frac {\partial \Gamma^\mu_{\nu\lambda}}{\partial x^\rho}
+\Gamma^\mu_{\lambda\sigma} \Gamma^\sigma_{\rho\nu}-\Gamma^\mu_{\rho\sigma}
\Gamma^\sigma_{\lambda\nu} \ .\label{B20}
\end{eqnarray}
(\ref{B20}) is a tensor field called the {\it curvature tensor}. Sometimes it is convenient to write it in the form
\begin{eqnarray}
{\RR^{\mu\nu}}_{\lambda\rho}=g^{\nu\alpha} {\RR^\mu}_{\alpha\lambda\rho} \ .\label{B21}
\end{eqnarray}
Moreover, one defines the {\it Ricci tensor} as
\begin{eqnarray}
\RR_{\mu\nu}:={\RR^\lambda}_{\mu\lambda\nu} \label{B22}
\end{eqnarray}
and the {\it curvature scalar}
\begin{eqnarray}
\RR:=g^{\mu\nu} \RR_{\mu\nu} \ .\label{B23}
\end{eqnarray}

%%%%%%%%%%%%%%%%%%%%%%%%%%%%%%%%%%%%%%%%%%%%%%%%%%%%%%%%%%%%%%%%%%%%%%%%%%%%%%%%%
\section{Killing equations} \label{app:Killing}
A symmetry (or isometry) of a manifold with metric is a transformation of the manifold which leaves the metric field invariant.
To understand what this means recall first that the metric $g_{\mu\nu}$, being a covariant tensor field, transforms as follows
under a change of coordinates $x^\mu\rightarrow \tilde x^\mu$
\begin{eqnarray}
\tilde g_{\mu\nu} (\tilde x) \frac {\partial \tilde x^\mu}{\partial x^\alpha} \frac {\partial \tilde
x^\nu}{\partial x^\beta}= g_{\alpha\beta} (x) \ . \label{B24}
\end{eqnarray}
If we interpret the map from the active point of view, it will define a transformation of the manifold: a point $p$ whose coordinates
are $x^\mu(p)=x^\mu$ is mapped to a new point $\tilde p $ whose coordinates (relative to the same coordinate patch) are $x^\mu(\tilde p)=\tilde x^\mu$.
Then, (\ref{B24}) establishes how the metric changes under such a transformation. We say that the map is an {\it isometry} if the metric
is left invariant under the transformation, namely if
\begin{eqnarray}
\tilde g_{\mu\nu} (\tilde x)= g_{\mu\nu} (\tilde x)\ . \label{B25}
\end{eqnarray}
If the transformation is an infinitesimal one, we can write it as $x^\mu\rightarrow \tilde x^\mu
=x^\mu+\varepsilon\xi^\mu(x)$, with $\varepsilon \ll 1$ and $\xi^\mu (x)$ is a contravariant vector field. For such an infinitesimal transformation,
inserting (\ref{B25}) into (\ref{B24}) and expanding to first order in $\varepsilon \xi^\mu$, gives
\begin{eqnarray}
0=\xi^\gamma \frac {\partial g_{\alpha\beta}}{\partial x^\gamma} +g_{\mu\beta} \frac {\partial \xi^\mu}{\partial x^\alpha} +g_{\alpha\mu}\frac {\partial
\xi^\mu}{\partial x^\beta} \ .\label{B26}
\end{eqnarray}
(\ref{B26}) can be rewritten in terms of the covariant components $\xi_\sigma =g_{\sigma\mu} \xi^\mu$ of the field $\xi$ as follows:
\begin{eqnarray}
0=\frac {\partial \xi_{\sigma}}{\partial x^\rho}+\frac {\partial \xi_{\rho}}{\partial x^\sigma}
+\xi_\lambda g^{\lambda\mu} \left(\frac {\partial g_{\alpha\beta}}{\partial x^\mu}-\frac {\partial g_{\mu\beta}}{\partial x^\alpha}
-\frac {\partial g_{\mu\alpha}}{\partial x^\beta} \right)\label{B27}
\end{eqnarray}
or, using (\ref{B6}) and (\ref{B12}),
\begin{eqnarray}
\nabla_\rho \xi_\sigma+\nabla_\sigma \xi_\rho=0 \ .\label{B28}
\end{eqnarray}
(\ref{B28}) are called the {\it Killing equations} and the fields
$\xi_\sigma$ which solve them, and which generate the isometries, are called the {\it Killing vectors} of the manifold.

%%%%%%%%%%%%%%%%%%%%%%%%%%%%%%%%%%%%
\subsection*{Killing vectors of the Minkowski hyperboloid}\label{hyperkilling}
We can easily find the Killing vectors of the Minkowski hyperboloid $dS(1,3)$, for any given value of the radius $R$. Indeed,
since $dS(1,3)$ is isometrically embedded in $M(1,4)$, its isometries are the isometries of $M(1,4)$ which leave each Minkowski hyperboloid,
solution of (\ref{hyperboloid}), invariant.
Now, the isometries of the flat metric $\eta_{AB}$ are the transformations of the Poincar\'e group acting on $M(1,4)$ whose subgroup which
stabilizes each $dS(1,3)$ is the de Sitter group $SO(1,4)$, namely the Lorentz group acting in one time plus four space dimensions.
The generators of $SO(1,4)$ are the vector fields (\ref{generators})
\begin{eqnarray}
N_{AB}= X_A \frac {\partial}{\partial X^B}-X_B \frac {\partial}{\partial X^A}  \label{B29}
\end{eqnarray}
and therefore the Killing fields of $dS(1,3)$ are simply the restrictions to $dS(1,3)$ itself of the fields (\ref{B29}).
Indeed, because the transformations of $SO(1,4)$ map $dS(1,3)$ onto itself, such restrictions are automatically tangent to it
and therefore define vector fields on $dS(1,3)$.
\end{appendix}

%%%%%%%%%%%%%%%%%%%%%%%%%%%%%%%%%%%%%%%%%%%%%%%%%%%%%%%%%%%%%%%%%%%%%%%%%%%%%%%%%%%%%%%%%%%%%%%%%%

\end{document}